# Ferro-rotational domain walls revealed by electric quadrupole second harmonic generation microscopy

Xiaoyu Guo[1], Rachel Owen[1], Austin Kaczmarek[1], Xiaochen Fang[2], Chandan De[3,4,5], Youngjun Ahn[1], Wei Hu[6,7], Nishkarsh Agarwal[8], Suk Hyun Sung[8], Robert Hovden[8], Sang-Wook Cheong[2], & Liuyan Zhao[1,+]

[1] *Department of Physics, University of Michigan, Ann Arbor, MI 48109, USA*

[2] *Rutgers Center for Emergent Materials and Department of Physics and Astronomy, Rutgers University, Piscataway, NJ 08854, USA*

[3] *Center for Artificial Low Dimensional Electronic Systems, Institute for Basic Science (IBS), Pohang 37673, Korea*

[4] *Laboratory of Pohang Emergent Materials, Pohang Accelerator Laboratory, Pohang 37673, Korea*

[5] *2D Crystal Consortium, Materials Research Institute, The Pennsylvania State University, University Park, PA 16802, USA*

[6] *School for Environment and Sustainability, University of Michigan, Ann Arbor, MI 48109, USA*

[7] *Department of Statistics, University of Michigan, Ann Arbor, MI 48109, USA*

[8] *Department of Material Science and Engineering, University of Michigan, Ann Arbor, MI 48109, USA*

[+] Corresponding to: lyzhao@umich.edu

**Abstract:** Domain walls are ubiquitous in materials that undergo phase transitions driven by spontaneous symmetry breaking. Domain walls in ferroics and multiferroics have received tremendous attention recently due to their emergent properties distinct from their domain counterparts, for example, their high mobility and controllability, as well as their potential applications in nanoelectronics. However, it is extremely challenging to detect, visualize and study the ferro-rotational (FR) domain walls because the FR order, in contrast to ferromagnetism (FM) and ferroelectricity (FE), is invariant under both the spatial-inversion and the time-reversal operations and thus hardly couple with conventional experimental probes. Here, an FR candidate $NiTiO_3$ is investigated by ultrasensitive electric quadrupole (EQ) second harmonic generation rotational anisotropy (SHG RA) to probe the point symmetries of the two degenerate FR domain states, showing their relation by the vertical mirror operations that are broken below the FR critical temperature. We then visualize the real-space FR domains by scanning EQ SHG microscopy, and further resolve the FR domain walls by revealing a suppressed SHG intensity at domain walls. By taking local EQ SHG RA measurements, we show the restoration of the mirror symmetry at FR domain walls and prove their unconventional nonpolar nature. Our findings not only provide a comprehensive insight into FR domain walls, but also demonstrate a unique and powerful tool for future studies on domain walls of unconventional ferroics, both of which pave the way towards future manipulations and applications of FR domain walls.

## Introduction

Domain walls are low-dimensional features that lie between neighboring domains related by broken symmetries through second order phase transitions in solids. In particular, domain walls in ferroics and multiferroics have attracted tremendous attention and have been intensively studied because of two perspectives. First, domain walls possess distinct symmetries from their domain counterparts and exhibit unique properties that are often absent inside the domains. For example, enhanced electrical conductivity (*1-5*), spontaneous polarizations (*6-11*) and exotic correlated magnetism (*12*) have been observed at the domain walls in various ferroic and multiferroic materials. More importantly, by coupling the ferroic orders to their conjugate fields, their domain walls can be controlled and manipulated, making the host materials promising for nanoelectronics applications such as diodes (*13*), nonvolatile memory (*14-16*), and tunnel junctions (*17*), where the domain walls play an active device part.

Ferro-rotational (FR) order, a ferroic order made of a toroidal arrangement of electric dipoles, distinguishes itself from other ferroic orders by its invariance under both the spatial-inversion and the time-reversal operations. Because of this unique symmetry property and its axial vector order parameter, there are no readily available conjugate vector fields that can couple to the FR order, making it an extremely challenging task to detect and distinguish the degenerate FR domains, as well as to control the domain walls between them. This FR order has been experimentally probed very recently (*18-21*), but its domain walls remain unexplored. Conventional techniques that have been used to investigate different types of ferroic domain walls, such as piezoresponse force microscopy (*22*), conductive atomic force microscopy (*1, 23*), nitrogen-vacancy center magnetometry (*24, 25*), magneto-optic Kerr effect (*26*) and electric dipole second harmonic generation (ED SHG) microscopy (*27-29*), barely couple to the FR order due to its unique symmetry properties and its axial order parameter characteristics.

Electric quadrupole (EQ) SHG, in contrast to ED SHG that only survives when spatial inversion in broken, is present in all systems while exhibiting much weaker intensities than ED SHG. EQ SHG can be described by $E_i(2\omega) \propto P_i^{eff}(2\omega) = \chi_{ijkl}^{EQ} E_j k_k E_l$, where $\chi_{ijkl}^{EQ}$ is the EQ SHG susceptibility tensor; $E_j$, $E_l$ and $E_i$ are electric fields of the incident and SHG light; and $k_k$ is the light wavevector. Comparing to linear optics, EQ SHG provides multiple copies of participating vector fields, $E_i(2\omega)$, $E_{j,l}(\omega)$, and $k_k$ that can combine to construct tensor fields, allowing a direct coupling with high-rank multipolar orders (*e.g.*, the FR order). Furthermore, the rank-4 EQ SHG susceptibility tensor $\chi_{ijkl}^{EQ}$ contains more tensor elements and is thus more sensitive to point symmetries than the linear optical counterpart.

In this study, based on its own merits, we apply EQ SHG to investigate the FR order of NiTiO$_3$ single crystals. By measuring the polarization rotational anisotropy (RA) of EQ SHG, we probe the point symmetries of the two degenerate FR domain states and reveal that those two states are related by vertical mirror operations. By performing scanning EQ SHG microscopy at the polarization angle with a maximum SHG intensity contrast between the two domain states, we manage to visualize the real-space FR domain structure in detail and quantify its physical properties. More importantly, by aligning the polarization angle at the vertical mirror direction where the two domain states have an equal SHG intensity, the FR domain walls are clearly visualized in the scanning microscopy as one-dimensional lines with a suppressed SHG intensity. We further confirm the presence of the mirror symmetry and the unconventional nonpolar nature at the domain walls with local EQ SHG RA measurements.

**Results and discussion**

We investigate the FR domains and domain walls on single crystal $NiTiO_3$. Above the FR phase transition temperature $T_c = 1560$ K, $NiTiO_3$ crystalizes in a corundum structure with the space group $R\bar{3}c$ (point group $\bar{3}m$) where the oxygen atoms form a distorted hexagonal close packing and two-thirds of the oxygen octahedra cages are randomly occupied by $Ni^{2+}$ and $Ti^{4+}$ (Fig. 1A). The structure possesses a $C_3$ rotational symmetry along the $c$-axis as well as three mirror planes parallel to the $c$-axis (Fig. 1B). Below $T_c$, ordering of the metallic ions forms where $Ni^{2+}$ and $Ti^{4+}$ are arranged alternatively in a layer-by-layer fashion (*30, 31*). The resulting two stacking sequences (Ni-Ti-Ni-Ti- and Ti-Ni-Ti-Ni-) give rise to two domain states A and B (Fig. 1C and 1E). These distinct domain states feature net structural rotations in opposite directions arising from the rotational distortions of the oxygen cages with different rotation directions and degrees of distortion between adjacent atomic layers (Fig. 1D and 1F). These distortions, which break the mirror symmetries and reduce the space group from $R\bar{3}c$ to $R\bar{3}$ (point group $\bar{3}$), result in a toroidal arrangement of electric dipole moments that lead to the FR order.

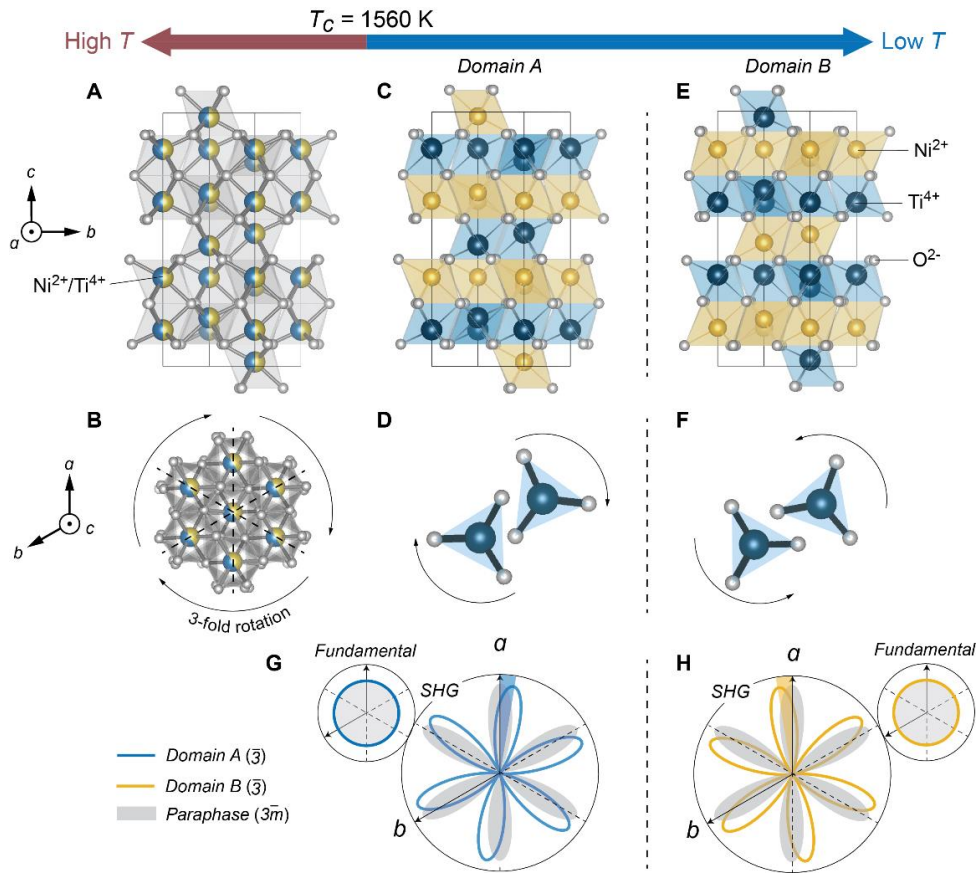

**FIG. 1. Crystal structures and FR domain states of NiTiO$_3$.** (**A**) *a*-axis view of the NiTiO$_3$ crystal unit cell above the FR phase transition temperature $T_c$. The bicolor spheres indicate either Ni$^{2+}$ or Ti$^{4+}$ while the white spheres represent O$^{2-}$. (**B**) *c*-axis view of the NiTiO$_3$ crystal structure above $T_c$. The 3-fold rotational symmetry and the three mirrors are indicated by the arrows and the dash lines, respectively. (**C** and **E**) *a*-axis view of the NiTiO$_3$ crystal unit cells of the FR domain state A and domain state B below $T_c$. The blue, yellow, and white spheres indicate Ti$^{4+}$, Ni$^{2+}$ and O$^{2-}$, respectively. (**D** and **F**) *c*-axis view of two oxygen cages enclosing Ti$^{4+}$ of domain A and B. Only two Ti$^{4+}$ and six O$^{2-}$ are depicted. Arrows indicate the rotation direction of the oxygen cages. (**G** and **H**) Simulated polarization-resolved fundamental reflection and the EQ SHG RA in the parallel channel from domain A and domain B under $\bar{3}$ point group (solid curves) and from the paraphase under $\bar{3}m$ point group (grey shaded areas).

Due to the absence of the leading order ED SHG for this centrosymmetric crystal structure, EQ SHG is the lowest rank nonlinear optical process that can couple with the FR order. Here, we simulate the EQ SHG RA results and show that it can be an effective approach to distinguishing the two degenerate FR domain states in NiTiO$_3$. Figure 1G and 1H show the simulated EQ SHG RA together with the polarization-resolved fundamental reflection in the parallel channel from FR domains A and B obtained under normal incidence of light along the *c*-axis of the sample. The blue and yellow solid curves are simulations based on the $\bar{3}$ point group of the FR domains while the grey shaded areas are those for the $\bar{3}m$ point group of the paraphase (*i.e.,* the high-symmetry phase above $T_c$) [see Supplementary Materials Section 1]. Clearly, whereas the polarization-resolved fundamental reflections show identical circular patterns from both domains, the EQ SHG RA patterns exhibit a rotation of the six-lobe patterns in opposite directions between the two FR domains with respect to those of the paraphase, demonstrating the advantage of EQ SHG over linear optics in probing the FR order and distinguishing the two degenerate FR domain states.

Experimentally, we first perform the EQ SHG RA measurements [see Methods] on a normal-cut sample (Fig. 2A) whose surface normal is off from the crystal *c*-axis by $\theta_1 = 2 \pm 1°$, determined by the group theory analysis [see Supplementary Materials Section 1] and confirmed by the X-ray diffraction. Surface ED contribution to the SHG signal has been ruled out by performing oblique incidence SHG RA measurements on this sample [see Supplementary Materials Section 2]. We have surveyed multiple locations across the sample and only observe two types of EQ SHG RA patterns. Figure 2B and 2C show the corrected EQ SHG RA patterns under the normal incidence of light [see Supplementary Materials Section 3] with the data (markers) fitted by the simulated functional forms from the point group $\bar{3}$ of the FR domains (solid lines). Both sets of the EQ SHG RA patterns possess the three-fold rotational symmetry but rotate away from the paraphase mirrors in opposite directions by 17°, which is the evidence of the broken mirror symmetry in the FR phase. Remarkably, this 34° rotation angle difference between the EQ SHG RA patterns is much more significant than that of the tiny rotational distortions of the crystal structures between the two domains. This rotation of the SHG RA pattern depends on the relative amplitudes between the susceptibility tensor elements of the paraphase and those of the FR phase which are wavelength-dependent (*19*) (*i.e.,* $\Delta = \frac{1}{3}\arctan\frac{\chi_{yyzy}(\lambda)}{\chi_{xxzx}(\lambda)}$, see Supplementary Materials Section 1). For NiTiO$_3$, the optical gap (~ 3eV, or ~400 nm) (*32, 33*) is close to our SHG wavelength (400 nm), where the optical resonance condition leads to the large rotation of the SHG RA patterns, compared to the delicate crystal cage rotation. Such large rotations in the EQ SHG RA patterns have also been observed in other FR materials (*19, 21, 34*), corroborating the benefits of EQ SHG RA in distinguishing the degenerate FR domains. Next, in preparation for the domain imaging, we prepared another tilt-cut NiTiO$_3$ sample that was quenched to room temperature after being annealed at 1300 °C for 12 hours. Its cut surface is deliberately chosen to be perpendicular to one of the three mirrors of the paraphase, with its surface normal off from the crystal *c*-axis by $\theta_2 = 34 \pm 6°$ (Fig. 2D). As shown in Fig. 2E and 2F, comparing to the normal-cut sample data, there is an increase of anisotropy together with a reduction of symmetry in the EQ SHG RA patterns for this tilt-cut sample. The six lobes of the EQ SHG RA patterns with even intensities in the normal-cut sample are reduced to two large and two small lobes in the tilt-cut sample along with the three-fold rotational symmetry reduced to two-fold. EQ SHG RA patterns from the two domains, however, are still related by the vertical mirror operation that is aligned with the *a*-axis of the crystal.

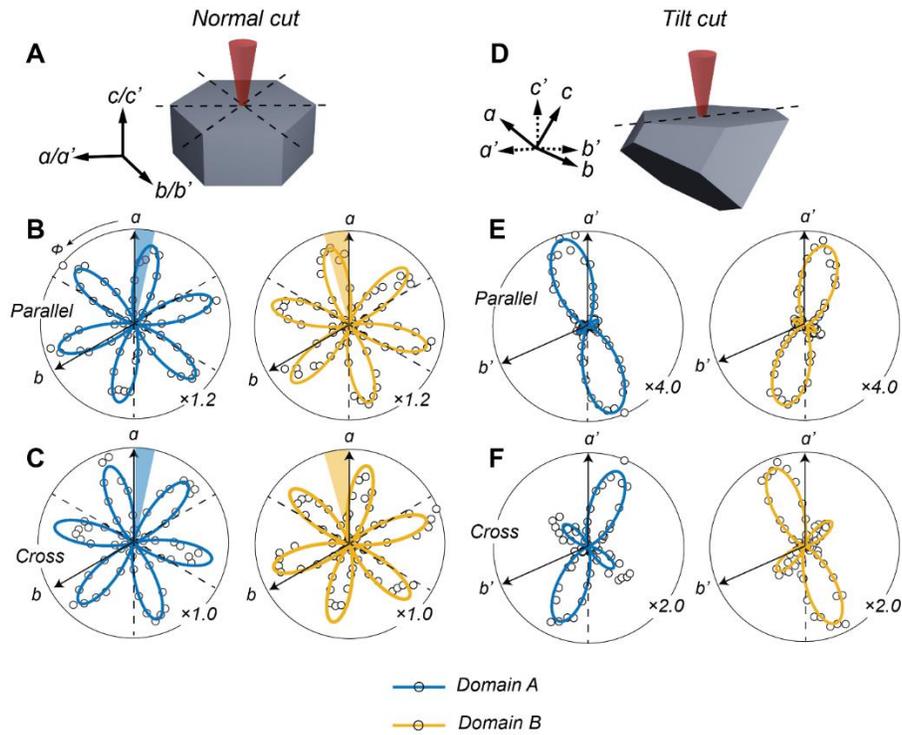

**FIG. 2. EQ SHG RA of the two FR domain states measured on the normal-cut and the tilt-cut samples.** (**A** and **D**) Schematics of the laser light incident on the normal-cut and the tilt-cut sample. Crystal axes are labelled where *a′*- and *b′*-axis are projections of *a*- and *b*-axis onto the sample cut surface and *c′*-axis is the projection of *c*-axis onto the surface normal. Mirrors of the paraphase are indicated by the dash lines. (**B** and **C**) Corrected EQ SHG RA patterns from the normal-cut sample FR domain A (blue) and domain B (yellow) in (**B**) the parallel and (**C**) the cross channel. Rotations of the patterns away from the *a*-axis are indicated by the colored shaded areas. (**E** and **F**) EQ SHG RA patterns from the tilt-cut sample FR domain A (blue) and domain B (yellow) in the (**E**) parallel and (**F**) the cross channel. Mirrors of the paraphase are indicated by the dash lines. Data (markers) are fitted by the functional form from group theory analysis (solid lines). Numbers at the bottom right of the plots indicate the scale of the plot.

The anisotropy in the tilt-cut sample introduces a large SHG intensity contrast in the parallel channel at the polarization angles $\phi_1 = -17°$ and $\phi_2 = 17°$ (insets of Fig. 3A and 3B) [see Supplementary Materials Section 4], allowing us to image the FR domains by the means of scanning EQ SHG microscopy [see Methods]. We first fix the polarization angle to be at $\phi_1 = 17°$ where domain A has a much larger intensity than that of domain B and perform the scanning EQ SHG measurement. The yielded FR domain map is shown in Fig. 3A [see Supplementary Materials Section 5 for color scale determination]. The two degenerate FR domain states are clearly distinguished with a stark contrast and are clearly separated. When the polarization is fixed at $\phi_2 = -17°$ where domain B has a larger SHG signal over domain A, a complimentary FR domain map is constructed and shown in Fig. 3B with the signal levels of the two FR domains flipped from Fig. 3A. Regions highlighted in yellow show exceptionally large SHG signals in both Fig. 3A and Fig. 3B. These regions correspond to the island-shaped NiO defects formed during sample growth [See Supplementary Materials Section 6].

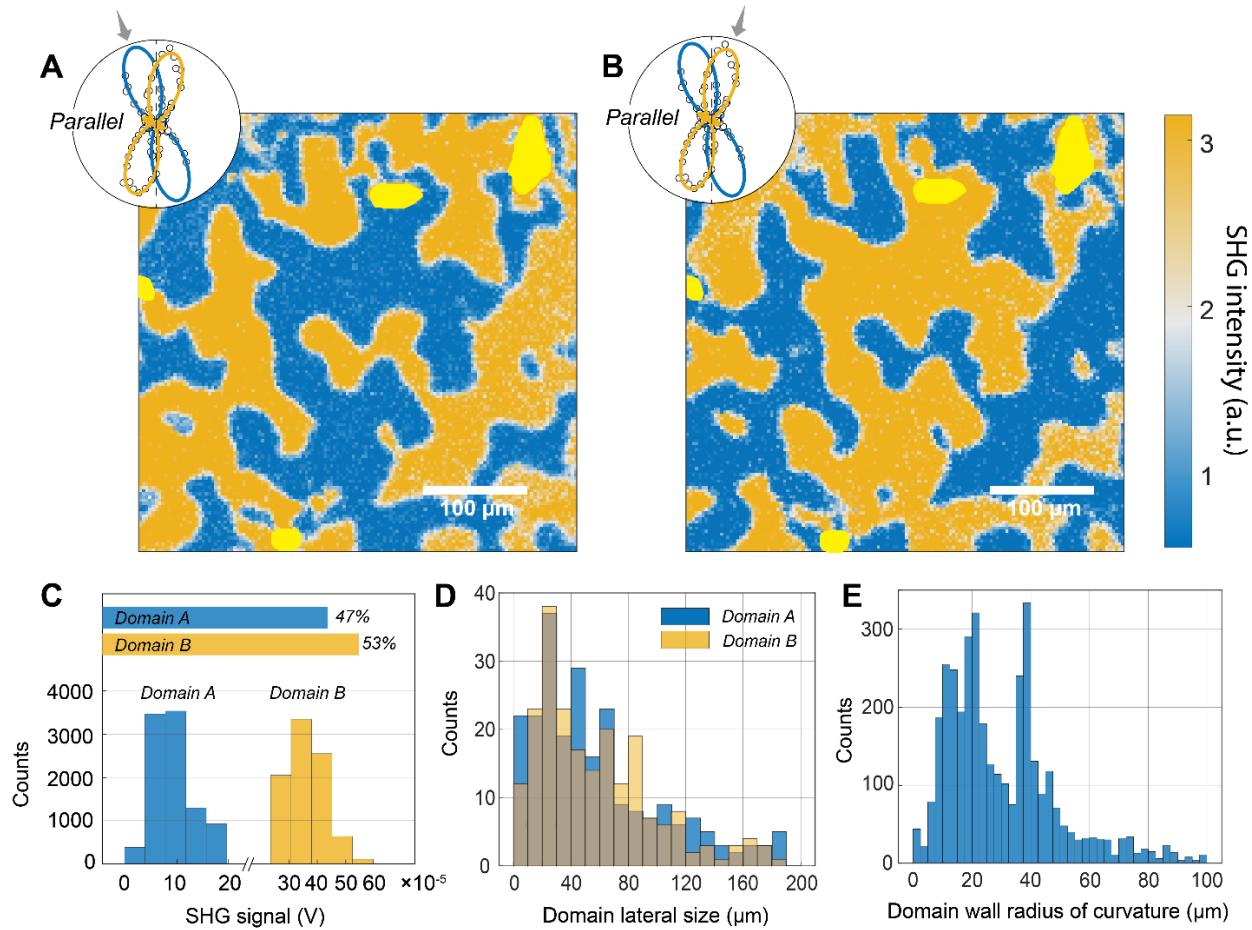

**FIG. 3. EQ SHG scanning microscopy of the FR domains.** (**A** and **B**) EQ SHG scanning of the FR domains in the parallel channel at the polarizations indicated by the arrows in the corresponding insets. The domain maps shown in (**A**) and (**B**) are flipped under the two selected polarizations. Regions with exceptionally large SHG signals are highlighted by the yellow color. (**C**) Histogram of the population of domain A and B. (**D**) Histogram of the domain lateral size of domain A and B. (**E**) Histogram of the radius of curvature along the domain walls.

The high-resolution FR domain imaging enables us to perform a series of statistical analysis on the FR domains and domain walls, which provide essential yet unprecedented information about the FR order and could be compared with ferroelectric (FE), ferromagnetic (FM) and the ferrotoroidal (FT) orders. To start with, we calculate the domain populations based on the SHG signal level of each pixel. As is shown in Fig. 3C, the populations of domain A and B are close to each other (47% and 53% for domain A and B, respectively), showing no obvious preference to either domain when the FR order is formed as the temperature cools down below $T_c$. We next investigate the lateral size of the FR domains obtained from the whole scanned region [see Methods] and plot the histogram of the lateral domain size in Fig. 3D. The peak frequency of the domain lateral size lies in 20-30 µm range, which is larger than the typical size of the FE domains that is less than one micron (*1, 23, 24, 35-40*), close to that of the FM domains that is of tens of micron (*41, 42*) and smaller than that of the FT domains that is of hundreds of micron (*29, 43*). During the sample preparation process, we observe that by quenching the sample across $T_c$, the domains can shrink by more than 80% of the size of the slow cooling sample domains [see Supplementary Materials Section 7], with an increase of the domain wall density. The cooling-rate dependence of the domain sizes has also been observed by previous studies (*20*). This indicates that the quadrupolar nature of the FR order is distinct from the dipolar ferroic counterparts such as FM and FE orders whose domain sizes mainly result from the energy competition between long-range (*e.g.*, magnetic dipole-dipole interaction for FM) and short-range (*e.g.*, exchange coupling for FM) interactions (*44*). On the contrary, instead of the static energetics, the dominant contributor to the formation of the FR domains and their size could be the non-equilibrium dynamics as well as the formation of topological defects during the cooling process of the sample growth (*45, 46*).

Besides the FR domains, thanks to the high resolution of the domain map, we can clearly visualize the boundaries between the two FR domains, *i.e.,* the domain walls. From the domain map, we observe that the domain walls twist and turn along the way. Figure 3E plots the histogram of the radius of curvature extracted from each point along the domain wall [see Methods and Supplementary Materials Section 8]. Two peak frequencies appear at 10-22.5 µm and 35-40 µm, which are comparable to the characteristic lateral domain sizes (20-30 µm). This is consistent with the meandering of the domain wall shown in Fig. 3A and 3B, and is in stark contrast to the straight wall morphology that is ubiquitous in, for example, strain-induced ferroelectric and ferroelastic materials (*1, 40, 47*). However, in FT (*29, 43*), FM (*41, 42*) and many of the improper FE materials (*12, 39, 48*), the domain walls also show curving profiles, similar to the case of FR domains in NiTiO$_3$. We infer from this observation that the domain walls tend to be curved when the order parameter is parallel to the out-of-plane rotational axis (*i.e.*, *c*-axis in NiTiO$_3$) and hence does not break any rotational symmetry. This is because such type of order parameters has zero component projected onto the in-plane domain walls and therefore is relatively insensitive to the curving of them.

Remarkably, scanning EQ SHG and local EQ SHG RA measurements can be applied to selectively image the domain walls and probe their symmetries. As shown in Fig. 4A, within the selected region shown in the lower inset of Fig. 4A, by choosing the polarization in the cross channel along the *a*-axis where both FR domains yield the same SHG signal level (Fig. 4A upper inset), the EQ SHG scanning image shows a clear suppression of SHG intensity at the domain walls as compared to the domains on both sides, highlighting the domain walls. The average SHG signal is calculated along the domain wall (green contour in Fig. 4B) as well as along the contours that are further away from it [see Methods]. The plot of the average SHG intensity as a function of the distance to the domain wall is then constructed and shown in Fig. 4C, with the error bars corresponding to the 95% confidence interval. The valley of the plot corresponds to the domain wall and shows a clear drop of the signal, as expected from the scanning EQ SHG image in Fig. 4A. Note that the SHG intensities from the two FR domains are similar but not identical. This is because the polarization is slightly off by less than 0.5° from the angle at which both FR domains yield the same SHG intensity. The drop of the signal at the boundaries is the signature of mirror symmetry along the incident light polarization, which forbids any SHG signal in the cross channel, in contrast to the absence of mirrors inside FR domains. We note that similar suppression of the SHG signal at the domain boundaries was also observed in polycrystalline $MoS_2$ (*49, 50*) due to the destructive interference between the SHG light from adjacent domains [discussed in Supplementary Materials Section 9 (*27, 29, 50, 51*)], in contrast to that observed on $NiTiO_3$ that results from the restoration of mirror symmetry on the domain walls. The uniform drop of the signal along the domain wall negates its polar nature since polar domain walls break the spatial-inversion symmetry and therefore are expected to show stronger SHG from the leading order ED contribution. For example, the 2 nm thick polar ferroelectric domain wall in $CaTiO_3$ can give rise to much more significant SHG signals than the nonpolar ferroelastic domains and are prominent in the scanning SHG mapping (*11*). The signal drop at the domain wall of our $NiTiO_3$ is $7.0 \pm 1.5\%$ compared to the average signal level of the two FR domains, yielding an estimated domain wall width to be $0.76 \pm 0.20$ μm [see Supplementary Materials Section 10]. This estimated domain wall width lies within the range of the previously reported values by STEM (*18*). Depending on the relative orientation of the domain walls to the surface normal, we may have overestimated the domain wall width by probing its tilt-cut cross section. Therefore, our measured wall width is an upper bound of the actual value.

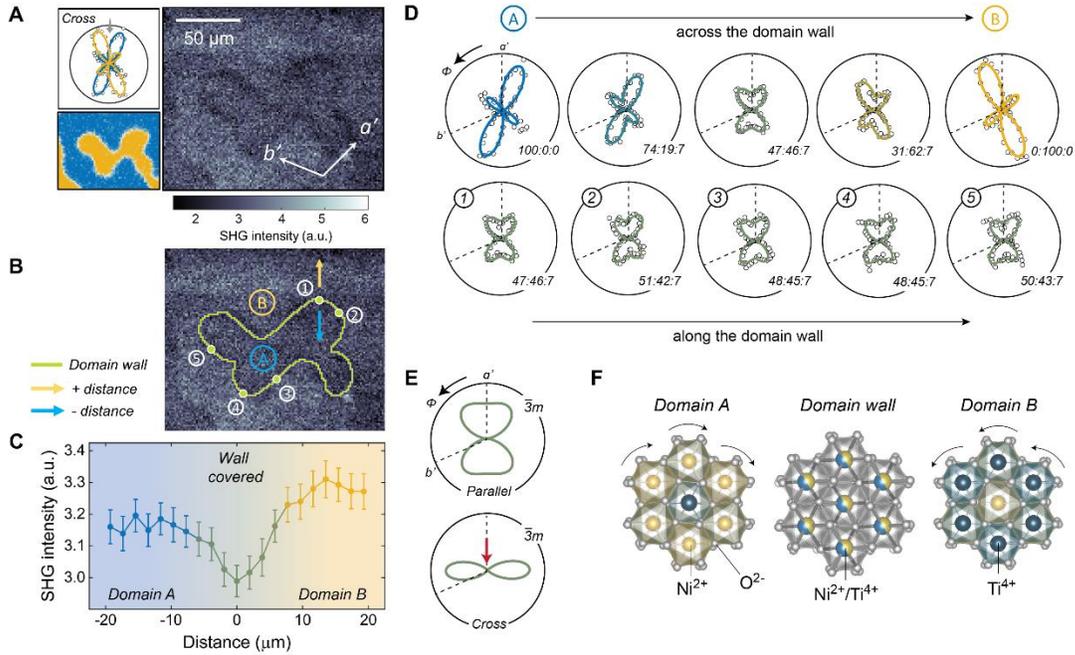

**FIG. 4. EQ SHG microscopy and RA of domain walls.** (**A**) EQ SHG scanning image of the domain walls in the cross channel at the polarization indicated by the arrow in the upper inset. The lower inset shows the EQ SHG scanning image of the domains extracted from Fig. 3A of the same region of interest. (**B**) The same EQ SHG scanning image of the domain walls as (**A**), where domain A and B are labelled. The green contour indicates the domain wall based on which (**C**) is constructed. The blue and yellow arrows indicate the positions where EQ SHG RA is measured and plotted in the upper row in (**D**). Numbers indicate the positions where EQ SHG RA is measured and plotted in the lower row in (**D**). (**C**) Plot of averaged EQ SHG intensity as a function of distance to the domain wall. (**D**) EQ SHG RA patterns in the cross channel measured across (upper row) and along (lower row) the domain wall. Experiment data (markers) are fitted by the functional form (solid curves) calculated from the point group $\bar{3}$. The numbers at the bottom right of the plots indicate the ratio of contributions from domain A, B, and the domain walls. All the EQ SHG RA are plotted on the same scale. (**E**) Simulated EQ SHG RA from a tilt-cut sample based on the point group $\bar{3}m$. The red arrow indicates the polarization parallel to the mirror where the cross channel shows zero SHG intensity. (**F**) c-axis view of the crystal structures of domain A, B, and the domain wall. Arrows indicate the rotation direction of the oxygen cages. The blue, yellow, and white spheres indicate $Ti^{4+}$, $Ni^{2+}$ and $O^{2-}$, respectively. The bicolor spheres indicate either $Ti^{4+}$ or $Ni^{2+}$.

To further characterize the domain walls, we perform a series of EQ SHG RA measurements across and along the domain wall. The upper row of Fig. 4D shows the EQ SHG RA measured across the domain wall along the blue and the yellow arrows shown in Fig. 4B. From domain A to domain B, there is a continuous evolution of the EQ SHG RA patterns and between the domains, the EQ SHG RA shows intermediate patterns which result from the superposition of the contributions from the two FR domains and the domain wall. Specifically, on the domain wall, the EQ SHG RA shows a bowtie shape which results from almost equal contributions from the two adjacent domains plus that from the domain wall. The lower row of Fig. 4D shows the EQ SHG RA patterns measured along the domain wall at the positions labelled by the numbers shown in Fig. 4B. All EQ SHG RA patterns measured at different positions on the domain wall show almost identical bowtie shapes. This confirms the nonpolar nature of the domain wall and shows stark contrast to the polar domain walls of ferroelastic materials such as $CaTiO_3$ (*11*) where the shapes of the SHG RA patterns are locked to the curving directions of domain walls.

Having established the presence of the mirror symmetry at domain walls, we here propose that the domain walls retain the crystal structure of the paraphase where $Ni^{2+}$ and $Ti^{4+}$ are distributed randomly and preserves the mirror symmetries, in contrast to the FR domains where the ordering of $Ni^{2+}$ and $Ti^{4+}$ takes place below $T_c$. To confirm this, Fig. 4E shows the simulated EQ SHG RA patterns from a tilt-cut sample based on the $\bar{3}m$ point group of the paraphase using the susceptibility tensor elements extracted from the fitting results of the FR domain data (Fig. 2E and 2F). The red arrow indicates that indeed, there is zero SHG signal present at the polarization aligned with the mirror in the cross channel, consistent with the intensity suppression at domain walls in Fig. 4A. We then construct the superposition of the EQ SHG RA patterns from both FR domains and the domain walls, which is used to fit the EQ SHG RA data in Fig. 4D (solid curves). The weight of the contributions from domain A, B and walls are displayed at the bottom right corner of each polar plot. Note that across the boundary from domain A to B, the contribution from domain A gradually decreases while that from domain B increases, whereas along the domain wall, the contributions from domain A and B are consistently almost equal. To summarize the structure change of the FR phase transition, as the temperature goes down, ordering of $Ni^{2+}$ and $Ti^{4+}$ takes place in a layer-by-layer fashion and form two degenerate FR domain states. At the intersection between the two domains, there is a region (the domain walls) where the crystal structure does not change and connect the two FR domains (Fig. 4F). Similar situations are also present in ferroelectricity such as $BiFeO_3$ where at the domain boundaries, it is more energetically favorable for the oxygen octahedra to remain unchanged (*52*).

**Conclusion**

In summary, we investigate the FR domains and domain walls in $NiTiO_3$ crystals by ultrasensitive EQ SHG techniques. The EQ SHG RA patterns from distinct FR domains rotate in the opposite directions from each other and are related by the vertical mirror operations that are present in the crystal, demonstrating the capability of EQ SHG RA to distinguish the degenerate FR domains. High-resolution FR domain maps are realized by scanning EQ SHG microscopy, showing almost-equal populations between the two degenerate domains with the domain lateral size mostly lying in the 20-30 μm range, much larger than the typical size of FE domains, close to that of the FM domains and smaller than that of the FT domains. More importantly, these EQ SHG techniques enable us to resolve the domain walls and probe their symmetries, despite the presence of spatial-inversion symmetry. We observe the meandering feature of the FR domain walls and confirm the presence of mirrors at the domain walls, as well as their nonpolar nature. The high-resolution FR domain and domain wall maps enabled by the EQ SHG techniques pave the way for the future study on the interactions between FR domain walls as well as the manipulation of FR domain walls by external stimuli. Moreover, it also provides a unique tool to study the interactions between the FR order and other orders coexisting in a system. It has been suggested that in many type-II multiferroics, the FR order is a prerequisite for subsequent noncentrosymmetric magnetic orders, which are the necessary ingredient for type-II multiferroicity (*53-55*). For NiTiO$_3$, it has been reported that an antiferromagnetic (AFM) order forms below $T_{\text{AFM}} = 23$ K (*56-58*). It would be insightful to explore how the AFM order is established upon the existing FR orders and how the AFM domains distribute in the presence of FR domains and domain walls.

**Materials and Methods:**

**Sample fabrication:** The single crystal of NiTiO$_3$ was grown by using a two-mirror type infrared image furnace in air after synthesizing polycrystalline feed rod by mixing NiO (99.999%) and TiO$_2$ (99.999%) in a stoichiometric ratio (*20, 59*). **Annealing history:** The tilt cut sample was annealed at 1300 °C for 12 hours and quenched to room temperature after growth. The normal cut sample was not annealed or quenched after growth. **Surface preparation:** The surfaces of both normal cut and tilt cut samples were mechanically polished on diamond lapping films with progressively finer grain size (30 μm to 0.1 μm). Final surface finishing was subsequently done on the mechanically polished surfaces using GIGA-0900 vibratory polisher with colloidal silica as the abrasive.

**EQ SHG measurements:** The 800 nm ultrafast laser source is of 50 fs pulse duration and 200 kHz repetition rate. It is focused down to a 15 μm diameter spot on the sample at a power of 3.7 mW, corresponding to a fluence of 10.5 mJ/cm$^2$. The laser light is shined normal to the sample surface and are moved across the sample surface to construct EQ SHG maps at various locations across the NiTiO$_3$ single crystal. For EQ SHG RA measurements, the same power and beam size are used. The polarizations of the incident and reflected light can be selected to be either parallel or perpendicular to each other, whose azimuthal angle $\phi$ can change correspondingly. The SHG signal is collected by a photomultiplier-tube, and then amplified by an electronic current pre-amplifier and a lock-in amplifier in sequence.

**Domain size analysis:** Multiple vertical and horizontal line-cuts have been extracted from the EQ SHG scanning maps shown in Fig. 3A and 3B. Lateral sizes of the domains are sampled from these line-cuts and are used to construct the domain lateral size histogram shown in Fig. 3D.

**Domain boundary analysis:** For the radius of curvature analysis, the domain walls are first picked out manually and then smoothened by the Polynomial Approximation with Exponential Kernel (PAEK) method based on a smoothing tolerance of 0.001 pixel size using ArcGIS software (*60*). The radius of curvature is then calculated at each point [see Supplementary Materials Section 8]. For the SHG signal analysis along the domain wall, the domain walls are first picked out manually by ArcGIS software. The distances from each pixel to the domain wall are then calculated by ArcGIS and the average SHG signals are calculated along each contour that has a fixed distance to the domain walls.

**Acknowledgements: Funding:** L. Z. acknowledges the support by NSF CAREER grant no. DMR-174774, AFOSR YIP grant no. FA9550-21-1-0065, and Alfred P. Sloan Foundation. The work at Rutgers University was supported by the DOE under Grant No. DOE: DE-FG02-07ER46382. R.H. acknowledges support from ARO grant no. W911NF-22-1-0056. **Author contributions:** X.G. performs the experiment and analyzes the data. R.O. analyzes the data. X.G. and A.K. constructs the SHG setup. C.D. grows the crystal under the guidance of S.C.. X.F. prepares the sample surface for the SHG measurements. W.H. analyzes the domain boundary data. N.A., S.H.S and R.H. perform the SEM measurements. X.G., Y.A. and L.Z. write the manuscript. **Completing interests:** The authors declare that they have no competing interests. **Data availability:** All data needed to evaluate the conclusions in the paper are present in the paper and the Supplementary Materials.

**References:**

1. J. Seidel, L. W. Martin, Q. He, Q. Zhan, Y. H. Chu, A. Rother, M. E. Hawkridge, P. Maksymovych, P. Yu, M. Gajek, N. Balke, S. V. Kalinin, S. Gemming, F. Wang, G. Catalan, J. F. Scott, N. A. Spaldin, J. Orenstein, R. Ramesh, Conduction at domain walls in oxide multiferroics. *Nat. Mater.* **8**, 229-234 (2009).
2. J. Guyonnet, I. Gaponenko, S. Gariglio, P. Paruch, Conduction at domain walls in insulating Pb(Zr$_{0.2}$Ti$_{0.8}$)O$_3$ thin films. *Adv. Mater.* **23**, 5377 (2011).
3. T. Sluka, A. K. Tagantsev, P. Bednyakov, N. Setter, Free-electron gas at charged domain walls in insulating BaTiO$_3$. *Nat. Commun.* **4**, 1808 (2013).
4. T. Rojac, A. Bencan, G. Drazic, N. Sakamoto, H. Ursic, B. Jancar, G. Tavcar, M. Makarovic, J. Walker, B. Malic, D. Damjanovic, Domain-wall conduction in ferroelectric BiFeO$_3$ controlled by accumulation of charged defects. *Nat. Mater.* **16**, 322 (2017).
5. J. Ma, J. Ma, Q. H. Zhang, R. C. Peng, J. Wang, C. Liu, M. Wang, N. Li, M. F. Chen, X. X. Cheng, P. Gao, L. Gu, L. Q. Chen, P. Yu, J. X. Zhang, C. W. Nan, Controllable conductive readout in self-assembled, topologically confined ferroelectric domain walls. *Nat. Nanotechnol.* **13**, 972-972 (2018).
6. H. Yokota, S. Matsumoto, E. K. H. Salje, Y. Uesu, Polar nature of domain boundaries in purely ferroelastic Pb$_3$(PO$_4$)$_2$ investigated by second harmonic generation microscopy. *Phys. Rev. B* **100**, 024101 (2019).
7. H. Yokota, S. Matsumoto, E. K. H. Salje, Y. Uesu, Symmetry and three-dimensional anisotropy of polar domain boundaries observed in ferroelastic LaAlO$_3$ in the complete absence of ferroelectric instability. *Phys. Rev. B* **98**, 104105 (2018).
8. X. K. Wei, A. K. Tagantsev, A. Kvasov, K. Roleder, C. L. Jia, N. Setter, Ferroelectric translational antiphase boundaries in nonpolar materials. *Nat. Commun.* **5**, 3031 (2014).
9. J. F. Scott, E. K. H. Salje, M. A. Carpenter, Domain wall damping and elastic softening in SrTiO$_3$: evidence for polar twin walls. *Phys. Rev. Lett.* **109**, 187601 (2012).
10. L. Goncalves-Ferreira, S. A. T. Redfern, E. Artacho, E. K. H. Salje, Ferrielectric twin walls in CaTiO$_3$. *Phys. Rev. Lett.* **101**, 097602 (2008).
11. H. Yokota, H. Usami, R. Haumont, P. Hicher, J. Kaneshiro, E. K. H. Salje, Y. Uesu, Direct evidence of polar nature of ferroelastic twin boundaries in CaTiO$_3$ obtained by second harmonic generation microscope. *Phys. Rev. B* **89**, 144109 (2014).
12. Y. N. Geng, N. Lee, Y. J. Choi, S. W. Cheong, W. D. Wu, Collective magnetism at multiferroic vortex domain walls. *Nano Lett.* **12**, 6055-6059 (2012).
13. J. R. Whyte, J. M. Gregg, A diode for ferroelectric domain-wall motion. *Nat. Commun.* **6**, 7361 (2015).
14. W. D. Yang, G. Tian, H. Fan, Y. Zhao, H. Y. Chen, L. Y. Zhang, Y. D. Wang, Z. Fan, Z. P. Hou, D. Y. Chen, J. W. Gao, M. Zeng, X. B. Lu, M. H. Qin, X. S. Gao, J. M. Liu, Nonvolatile ferroelectric-domain-wall memory embedded in a complex topological domain structure. *Adv. Mater.* **34**, 2107711 (2022).
15. H. W. Shin, J. H. Son, J. Y. Son, Current change due to artificial patterning of the number of ferroelectric domain walls and nonvolatile memory characteristics. *Appl. Phys. Lett.* **119**, 122901 (2021).
16. P. Sharma, Q. Zhang, D. Sando, C. H. Lei, Y. Y. Liu, J. Y. Li, V. Nagarajan, J. Seidel, Nonvolatile ferroelectric domain wall memory. *Sci. Adv.* **3**, e1700512 (2017).
17. G. Sanchez-Santolino, J. Tornos, D. Hernandez-Martin, J. I. Beltran, C. Munuera, M. Cabero, A. Perez-Munoz, J. Ricote, F. Mompean, M. Garcia-Hernandez, Z. Sefrioui, C. Leon, S. J. Pennycook,

M. C. Munoz, M. Varela, J. Santamaria, Resonant electron tunnelling assisted by charged domain walls in multiferroic tunnel junctions. *Nat. Nanotechnol.* **12**, 655 (2017).
18. T. Hayashida, Y. Uemura, K. Kimura, S. Matsuoka, D. Morikawa, S. Hirose, K. Tsuda, T. Hasegawa, T. Kimura, Visualization of ferroaxial domains in an order-disorder type ferroaxial crystal. *Nat. Commun.* **11**, 4582 (2020).
19. W. C. Jin, E. Drueke, S. W. Li, A. Admasu, R. Owen, M. Day, K. Sun, S. W. Cheong, L. Y. Zhao, Observation of a ferro-rotational order coupled with second-order nonlinear optical fields. *Nat. Phys.* **16**, 42 (2020).
20. T. Hayashida, Y. Uemura, K. Kimura, S. Matsuoka, M. Hagihala, S. Hirose, H. Morioka, T. Hasegawa, T. Kimura, Phase transition and domain formation in ferroaxial crystals. *Phys. Rev. Mater.* **5**, 124409 (2021).
21. X. P. Luo, D. Obeysekera, C. Won, S. H. Sung, N. Schnitzer, R. Hovden, S. W. Cheong, J. J. Yang, K. Sun, L. Y. Zhao, Ultrafast modulations and detection of a ferro-rotational charge density wave using time-resolved electric quadrupole second harmonic generation. *Phys. Rev. Lett.* **127**, 126401 (2021).
22. A. Gruverman, M. Alexe, D. Meier, Piezoresponse force microscopy and nanoferroic phenomena. *Nat. Commun.* **10**, 1661 (2019).
23. D. Meier, J. Seidel, A. Cano, K. Delaney, Y. Kumagai, M. Mostovoy, N. A. Spaldin, R. Ramesh, M. Fiebig, Anisotropic conductance at improper ferroelectric domain walls. *Nat. Mater.* **11**, 284-288 (2012).
24. I. Gross, W. Akhtar, V. Garcia, L. J. Martinez, S. Chouaieb, K. Garcia, C. Carretero, A. Barthelemy, P. Appel, P. Maletinsky, J. V. Kim, J. Y. Chauleau, N. Jaouen, M. Viret, M. Bibes, S. Fusil, V. Jacques, Real-space imaging of non-collinear antiferromagnetic order with a single-spin magnetometer. *Nature* **549**, 252 (2017).
25. F. Casola, T. van der Sar, A. Yacoby, Probing condensed matter physics with magnetometry based on nitrogen-vacancy centres in diamond. *Nat. Rev. Mater.* **3**, 17088 (2018).
26. Z. C. Luo, A. Hrabec, T. P. Dao, G. Sala, S. Finizio, J. X. Feng, S. Mayr, J. Raabe, P. Gambardella, L. J. Heyderman, Current-driven magnetic domain-wall logic. *Nature* **579**, 214 (2020).
27. M. Trassin, G. De Luca, S. Manz, M. Fiebig, Probing ferroelectric domain engineering in $BiFeO_3$ thin films by second harmonic generation. *Adv. Mater.* **27**, 4871-4876 (2015).
28. D. Meier, M. Maringer, T. Lottermoser, P. Becker, L. Bohaty, M. Fiebig, Observation and coupling of domains in a spin-spiral multiferroic. *Phys. Rev. Lett.* **102**, 107202 (2009).
29. B. B. Van Aken, J. P. Rivera, H. Schmid, M. Fiebig, Observation of ferrotoroidic domains. *Nature* **449**, 702-705 (2007).
30. H. Boysen, F. Frey, M. Lerch, T. Vogt, A neutron powder investigation of the high-temperature phase-transition in $NiTiO_3$. *Z Kristallogr Cryst Mater* **210**, 328-337 (1995).
31. M. Lerch, H. Boysen, R. Neder, F. Frey, W. Laqua, Neutron-scattering investigation of the high-temperature phase-transition in $NiTiO_3$. *J. Phys. Chem. Solids* **53**, 1153-1156 (1992).
32. M. W. Li, J. P. Yuan, X. M. Gao, E. Q. Liang, C. Y. Wang, Structure and optical absorption properties of $NiTiO_3$ nanocrystallites. *Appl. Phys. A Mater. Sci. Process* **122**, (2016).
33. Y. J. Lin, Y. H. Chang, W. D. Yang, B. S. Tsai, Synthesis and characterization of ilmenite $NiTiO_3$ and $CoTiO_3$ prepared by a modified Pechini method. *J. Non Cryst. Solids* **352**, 789-794 (2006).
34. R. Owen, E. Drueke, C. Albunio, A. Kaczmarek, W. C. Jin, D. Obeysekera, S. W. Cheong, J. J. Yang, S. Cundiff, L. Y. Zhao, Second-order nonlinear optical and linear ultraviolet-visible absorption properties of the type-II multiferroic candidates $RbFe(AO_4)_2$ (A = Mo, Se, S). *Phys. Rev. B* **103**, 054104 (2021).
35. A. Schilling, R. M. Bowman, G. Catalan, J. F. Scott, J. M. Gregg, Morphological control of polar orientation in single-crystal ferroelectric nanowires. *Nano Lett.* **7**, 3787-3791 (2007).

36. C. J. M. Daumont, D. Mannix, S. Venkatesan, G. Catalan, D. Rubi, B. J. Kooi, J. T. M. De Hosson, B. Noheda, Epitaxial TbMnO$_3$ thin films on SrTiO$_3$ substrates: a structural study. *J. Phys. Condens. Matter.* **21**, 182001 (2009).
37. C. C. Neacsu, B. B. van Aken, M. Fiebig, M. B. Raschke, Second-harmonic near-field imaging of ferroelectric domain structure of YMnO$_3$. *Phys. Rev. B* **79**, 100107(R) (2009).
38. A. Schilling, D. Byrne, G. Catalan, K. G. Webber, Y. A. Genenko, G. S. Wu, J. F. Scott, J. M. Gregg, Domains in ferroelectric nanodots. *Nano Lett.* **9**, 3359-3364 (2009).
39. T. Choi, Y. Horibe, H. T. Yi, Y. J. Choi, W. D. Wu, S. W. Cheong, Insulating interlocked ferroelectric and structural antiphase domain walls in multiferroic YMnO$_3$. *Nat. Mater.* **9**, 253-258 (2010).
40. J. Seidel, P. Maksymovych, Y. Batra, A. Katan, S. Y. Yang, Q. He, A. P. Baddorf, S. V. Kalinin, C. H. Yang, J. C. Yang, Y. H. Chu, E. K. H. Salje, H. Wormeester, M. Salmeron, R. Ramesh, Domain wall conductivity in La-doped BiFeO$_3$. *Phys. Rev. Lett.* **105**, 197603 (2010).
41. W. Kuch, "Imaging magnetic microspectroscopy" in *Magnetic Microscopy of Nanostructures,* H. Hopster, H. P. Oepen, Eds. (Springer Berlin Heidelberg, Berlin, Heidelberg, 2005), pp. 1-28.
42. W. Szmaja, Recent developments in the imaging of magnetic domains. *Adv. Imaging. Electron. Phys.* **141**, 175-256 (2006).
43. A. S. Zimmermann, D. Meier, M. Fiebig, Ferroic nature of magnetic toroidal order. *Nat. Commun.* **5**, 4796 (2014).
44. D. M. Evans, V. Garcia, D. Meier, M. Bibes, Domains and domain walls in multiferroics. *Physical Sciences Reviews* **5**, 20190067 (2020).
45. H. Okino, J. Sakamoto, T. Yamamoto, Cooling-rate-dependent domain structures of Pb(Mg$_{1/3}$Nb$_{2/3}$)O$_3$-PbTiO$_3$ single crystals observed by contact resonance piezoresponse force microscopy. *Jpn. J. Appl. Phys.* **43**, 6808-6811 (2004).
46. S. C. Chae, N. Lee, Y. Horibe, M. Tanimura, S. Mori, B. Gao, S. Carr, S. W. Cheong, Direct observation of the proliferation of ferroelectric loop domains and vortex-antivortex pairs. *Phys. Rev. Lett.* **108**, 167603 (2012).
47. Y. Uesu, S. Kurimura, Y. Yamamoto, Optical second harmonic images of 90° domain structure in BaTiO$_3$ and periodically inverted antiparallel domains in LiTaO$_3$. *Appl. Phys. Lett.* **66**, 2165 (1995).
48. T. Jungk, A. Hoffmann, M. Fiebig, E. Soergel, Electrostatic topology of ferroelectric domains in YMnO$_3$. *Appl. Phys. Lett.* **97**, 012904 (2010).
49. J. X. Cheng, T. Jiang, Q. Q. Ji, Y. Zhang, Z. M. Li, Y. W. Shan, Y. F. Zhang, X. G. Gong, W. T. Liu, S. W. Wu, Kinetic nature of grain boundary formation in as-grown MoS$_2$ monolayers. *Adv. Mater.* **27**, 4069-4074 (2015).
50. X. B. Yin, Z. L. Ye, D. A. Chenet, Y. Ye, K. O'Brien, J. C. Hone, X. Zhang, Edge nonlinear optics on a MoS$_2$ atomic monolayer. *Science* **344**, 488-490 (2014).
51. N. Leo, A. Bergman, A. Cano, N. Poudel, B. Lorenz, M. Fiebig, D. Meier, Polarization control at spin-driven ferroelectric domain walls. *Nat. Commun.* **6**, 6661 (2015).
52. A. Lubk, S. Gemming, N. A. Spaldin, First-principles study of ferroelectric domain walls in multiferroic bismuth ferrite. *Phys. Rev. B* **80**, 104110 (2009).
53. R. D. Johnson, S. Nair, L. C. Chapon, A. Bombardi, C. Vecchini, D. Prabhakaran, A. T. Boothroyd, P. G. Radaelli, Cu$_3$Nb$_2$O$_8$: a multiferroic with chiral coupling to the crystal structure. *Phys. Rev. Lett.* **107**, 137205 (2011).
54. A. J. Hearmon, F. Fabrizi, L. C. Chapon, R. D. Johnson, D. Prabhakaran, S. V. Streltsov, P. J. Brown, P. G. Radaelli, Electric field control of the magnetic chiralities in ferroaxial multiferroic RbFe(MoO$_4$)$_2$. *Phys. Rev. Lett.* **108**, 237201 (2012).
55. R. D. Johnson, L. C. Chapon, D. D. Khalyavin, P. Manuel, P. G. Radaelli, C. Martin, Giant improper ferroelectricity in the ferroaxial magnet CaMn$_7$O$_{12}$. *Phys. Rev. Lett.* **108**, 067201 (2012).

56. G. Shirane, S. J. Pickart, Y. Ishikawa, Neutron diffraction study of antiferromagnetic $MnTiO_3$ and $NiTiO_3$. *J. Phys. Soc. Jpn.* **14**, 1352-1360 (1959).
57. A. Ito, H. Kawano, H. Yoshizawa, K. Motoya, Magnetic-properties and phase-diagram of $Ni_xMn_{1-x}TiO_3$. *J. Magn. Magn. Mater.* **104**, 1637-1638 (1992).
58. Y. Yamaguchi, T. Nakano, Y. Nozue, T. Kimura, Magnetoelectric effect in an XY-like spin glass system $Ni_xMn_{1-x}TiO_3$. *Phys. Rev. Lett.* **108**, 057203 (2012).
59. K. Dey, S. Sauerland, J. Werner, Y. Skourski, M. Abdel-Hafiez, R. Bag, S. Singh, R. Klingeler, Magnetic phase diagram and magnetoelastic coupling of $NiTiO_3$. *Phys. Rev. B* **101**, 195122 (2020).
60. E. Bodansky, A. Gribov, M. Pilouk, Smoothing and compression of lines obtained by raster-to-vector conversion. *Graphics Recognition: Algorithms and Applications* **2390**, 256-265 (2002).

# Supplementary Materials for

## Ferro-rotational domain walls revealed by electric quadrupole second harmonic generation microscopy

Xiaoyu Guo[1], Rachel Owen[1], Austin Kaczmarek[1], Xiaochen Fang[2], Chandan De[3,4,5], Youngjun Ahn[1], Wei Hu[6,7], Nishkarsh Agarwal[8], Suk Hyun Sung[8], Robert Hovden[8], Sang-Wook Cheong[2], & Liuyan Zhao[1,+]

[1] *Department of Physics, University of Michigan, Ann Arbor, MI 48109, USA*

[2] *Rutgers Center for Emergent Materials and Department of Physics and Astronomy, Rutgers University, Piscataway, NJ 08854, USA*

[3] *Center for Artificial Low Dimensional Electronic Systems, Institute for Basic Science (IBS), Pohang 37673, Korea*

[4] *Laboratory of Pohang Emergent Materials, Pohang Accelerator Laboratory, Pohang 37673, Korea*

[5] *2D Crystal Consortium, Materials Research Institute, The Pennsylvania State University, University Park, PA 16802, USA*

[6] *School for Environment and Sustainability, University of Michigan, Ann Arbor, MI 48109, USA*

[7] *Department of Statistics, University of Michigan, Ann Arbor, MI 48109, USA*

[8] *Department of Material Science and Engineering, University of Michigan, Ann Arbor, MI 48109, USA*

[+] Corresponding to: lyzhao@umich.edu

**Table of Contents**

1. **Group theory analysis of EQ SHG RA**
2. **Oblique SHG RA measurement on the unannealed normal-cut NiTiO$_3$**
3. **Normal incidence EQ SHG RA of the unannealed normal-cut NiTiO$_3$**
4. **SHG signal contrast between the normal-cut and the tilt-cut sample**
5. **FR domain maps color scale determination**
6. **Exceptionally large SHG signals from NiO defects**
7. **EQ SHG scanning microscopy of the unannealed normal-cut NiTiO$_3$**
8. **Radius of curvature along the domain walls of the annealed and quenched tilt-cut NiTiO$_3$**
9. **Discussion of destructive interference at the domain boundary**
10. **Estimation of the domain wall width**

# 1. Group theory analysis of EQ SHG RA

### a. On the normal-cut sample with FR order ($\bar{3}$)

For domain A, the simulated EQ SHG intensity as a function of susceptibility tensors and the azimuth angle $\phi$ of the input polarizations are:

$$SHG_{parallel,A} = (\chi_{xxzx} \cos 3\phi - \chi_{yyzy} \sin 3\phi)^2$$
$$SHG_{cross,A} = (\chi_{yyzy} \cos 3\phi + \chi_{xxzx} \sin 3\phi)^2.$$

For domain B:

$$SHG_{parallel,B} = (\chi_{xxzx} \cos 3\phi + \chi_{yyzy} \sin 3\phi)^2$$
$$SHG_{cross,B} = (\chi_{yyzy} \cos 3\phi - \chi_{xxzx} \sin 3\phi)^2$$

### b. On the normal-cut sample in the paraphase ($\bar{3}m$)

$$SHG_{parallel} = (\chi_{xxzx} \cos 3\phi)^2$$
$$SHG_{cross} = (\chi_{xxzx} \sin 3\phi)^2$$

### c. On the tilt-cut sample with FR order ($\bar{3}$)

For domain A:

$$\begin{aligned}
SHG_{parallel,A} = &(\sin[\phi](-\text{Sin}[\phi](\text{Cos}[\phi](-\text{Cos}[\beta]^2(\chi_{xxzx}\text{Cos}[3\alpha] + \chi_{yyzy}\text{Sin}[3\alpha]) \\
&+ (-\chi_{xyyx} + \chi_{yzzy})\text{Cos}[\beta]\text{Sin}[\beta] + (\chi_{xzxx}\text{Cos}[3\alpha] + \chi_{yzyy}\text{Sin}[3\alpha])\text{Sin}[\beta]^2) \\
&- (\chi_{yyzy}\text{Cos}[3\alpha]\text{Cos}[\beta] - \chi_{xxzx}\text{Cos}[\beta]\text{Sin}[3\alpha] - \chi_{yyxy}\text{Sin}[\beta])\text{Sin}[\phi]) \\
&+ \text{Cos}[\phi](\text{Cos}[\phi](\chi_{xxzx}\text{Cos}[\beta]^3\text{Sin}[3\alpha] - \frac{1}{2}(\chi_{xyyy} - 2\chi_{xzzy} \\
&+ \chi_{yyxy})\text{Cos}[\beta]^2\text{Sin}[\beta] - (\chi_{xzxx} + \chi_{zxxx})\text{Cos}[\beta]\text{Sin}[3\alpha]\text{Sin}[\beta]^2 \\
&+ \chi_{zxyz}\text{Sin}[\beta]^3 + \text{Cos}[3\alpha]\text{Cos}[\beta](-\chi_{yyzy}\text{Cos}[\beta]^2 + (\chi_{yzyy} \\
&+ \chi_{zyyy})\text{Sin}[\beta]^2)) - (-\text{Cos}[\beta]^2(\chi_{xxzx}\text{Cos}[3\alpha] + \chi_{yyzy}\text{Sin}[3\alpha]) + (-\chi_{xyxy} \\
&+ \chi_{zyzy})\text{Cos}[\beta]\text{Sin}[\beta] + (\chi_{zxxx}\text{Cos}[3\alpha] + \chi_{zyyy}\text{Sin}[3\alpha])\text{Sin}[\beta]^2)\text{Sin}[\phi])) \\
&- \text{Cos}[\phi](-\text{Sin}[\phi](\text{Cos}[\phi](\chi_{xxzx}\text{Cos}[\beta]^3\text{Sin}[3\alpha] + (\chi_{xyyy} \\
&- 2\chi_{xzzy})\text{Cos}[\beta]^2\text{Sin}[\beta] - 2\chi_{xzxx}\text{Cos}[\beta]\text{Sin}[3\alpha]\text{Sin}[\beta]^2 + \chi_{xzyz}\text{Sin}[\beta]^3 \\
&+ \text{Cos}[3\alpha](-\chi_{yyzy}\text{Cos}[\beta]^3 + \chi_{yzyy}\text{Sin}[\beta]\text{Sin}[2\beta])) \\
&- (-\text{Cos}[\beta]^2(\chi_{xxzx}\text{Cos}[3\alpha] + \chi_{yyzy}\text{Sin}[3\alpha]) + (-\chi_{xyyx} \\
&+ \chi_{yzzy})\text{Cos}[\beta]\text{Sin}[\beta] + (\chi_{xzxx}\text{Cos}[3\alpha] + \chi_{yzyy}\text{Sin}[3\alpha])\text{Sin}[\beta]^2)\text{Sin}[\phi]) \\
&+ \text{Cos}[\phi](\text{Cos}[\beta]\text{Cos}[\phi](\chi_{yyzy}\text{Cos}[\beta]^3\text{Sin}[3\alpha] + (-\chi_{xyxy} - 2\chi_{xyyx} + 2\chi_{yzzy} \\
&+ \chi_{zyzy})\text{Cos}[\beta]^2\text{Sin}[\beta] - (2\chi_{yzyy} + \chi_{zyyy})\text{Cos}[\beta]\text{Sin}[3\alpha]\text{Sin}[\beta]^2 + (-\chi_{yzyz} \\
&- 2\chi_{zyyz} + \chi_{zzzz})\text{Sin}[\beta]^3 + \text{Cos}[3\alpha]\text{Cos}[\beta](\chi_{xxzx}\text{Cos}[\beta]^2 - (2\chi_{xzxx} \\
&+ \chi_{zxxx})\text{Sin}[\beta]^2)) - (\chi_{xxzx}\text{Cos}[\beta]^3\text{Sin}[3\alpha] - \frac{1}{2}(\chi_{xyyy} - 2\chi_{xzzy} \\
&+ \chi_{yyxy})\text{Cos}[\beta]^2\text{Sin}[\beta] - (\chi_{xzxx} + \chi_{zxxx})\text{Cos}[\beta]\text{Sin}[3\alpha]\text{Sin}[\beta]^2 \\
&+ \chi_{zxyz}\text{Sin}[\beta]^3 + \text{Cos}[3\alpha]\text{Cos}[\beta](-\chi_{yyzy}\text{Cos}[\beta]^2 + (\chi_{yzyy} \\
&+ \chi_{zyyy})\text{Sin}[\beta]^2))\text{Sin}[\phi])))^2;
\end{aligned}$$

$$\begin{aligned}SHG_{cross,A} = &(\text{Sin}[\phi](\text{Cos}[\phi](\text{Cos}[\phi](-\text{Cos}[\beta]^2(\chi_{xxzx}\text{Cos}[3\alpha]+\chi_{yyzy}\text{Sin}[3\alpha])+(-\chi_{xyyx}\\&+\chi_{yzzy})\text{Cos}[\beta]\text{Sin}[\beta]+(\chi_{xzxx}\text{Cos}[3\alpha]+\chi_{yzyy}\text{Sin}[3\alpha])\text{Sin}[\beta]^2)\\&-(\chi_{yyzy}\text{Cos}[3\alpha]\text{Cos}[\beta]-\chi_{xxzx}\text{Cos}[\beta]\text{Sin}[3\alpha]-\chi_{yyxy}\text{Sin}[\beta])\text{Sin}[\phi])\\&+\text{Sin}[\phi](\text{Cos}[\phi](\chi_{xxzx}\text{Cos}[\beta]^3\text{Sin}[3\alpha]-\tfrac{1}{2}(\chi_{xyyy}-2\chi_{xzzy}\\&+\chi_{yyxy})\text{Cos}[\beta]^2\text{Sin}[\beta]-(\chi_{xzxx}+\chi_{zxxx})\text{Cos}[\beta]\text{Sin}[3\alpha]\text{Sin}[\beta]^2\\&+\chi_{zxyz}\text{Sin}[\beta]^3+\text{Cos}[3\alpha]\text{Cos}[\beta](-\chi_{yyzy}\text{Cos}[\beta]^2+(\chi_{yzyy}\\&+\chi_{zyyy})\text{Sin}[\beta]^2))-(-\text{Cos}[\beta]^2(\chi_{xxzx}\text{Cos}[3\alpha]+\chi_{yyzy}\text{Sin}[3\alpha])+(-\chi_{xyxy}\\&+\chi_{zyzy})\text{Cos}[\beta]\text{Sin}[\beta]+(\chi_{zxxx}\text{Cos}[3\alpha]+\chi_{zyyy}\text{Sin}[3\alpha])\text{Sin}[\beta]^2)\text{Sin}[\phi]))\\&-\text{Cos}[\phi](\text{Cos}[\phi](\text{Cos}[\phi](\chi_{xxzx}\text{Cos}[\beta]^3\text{Sin}[3\alpha]+(\chi_{xyyy}\\&-2\chi_{xzzy})\text{Cos}[\beta]^2\text{Sin}[\beta]-2\chi_{xzxx}\text{Cos}[\beta]\text{Sin}[3\alpha]\text{Sin}[\beta]^2+\chi_{xzyz}\text{Sin}[\beta]^3\\&+\text{Cos}[3\alpha](-\chi_{yyzy}\text{Cos}[\beta]^3+\chi_{yzyy}\text{Sin}[\beta]\text{Sin}[2\beta]))\\&-(-\text{Cos}[\beta]^2(\chi_{xxzx}\text{Cos}[3\alpha]+\chi_{yyzy}\text{Sin}[3\alpha])+(-\chi_{xyyx}\\&+\chi_{yzzy})\text{Cos}[\beta]\text{Sin}[\beta]+(\chi_{xzxx}\text{Cos}[3\alpha]+\chi_{yzyy}\text{Sin}[3\alpha])\text{Sin}[\beta]^2)\text{Sin}[\phi])\\&+\text{Sin}[\phi](\text{Cos}[\beta]\text{Cos}[\phi](\chi_{yyzy}\text{Cos}[\beta]^3\text{Sin}[3\alpha]+(-\chi_{xyxy}-2\chi_{xyyx}+2\chi_{yzzy}\\&+\chi_{zyzy})\text{Cos}[\beta]^2\text{Sin}[\beta]-(2\chi_{yzyy}+\chi_{zyyy})\text{Cos}[\beta]\text{Sin}[3\alpha]\text{Sin}[\beta]^2+(-\chi_{yzyz}\\&-2\chi_{zyyz}+\chi_{zzzz})\text{Sin}[\beta]^3+\text{Cos}[3\alpha]\text{Cos}[\beta](\chi_{xxzx}\text{Cos}[\beta]^2-(2\chi_{xzxx}\\&+\chi_{zxxx})\text{Sin}[\beta]^2))-(\chi_{xxzx}\text{Cos}[\beta]^3\text{Sin}[3\alpha]-\tfrac{1}{2}(\chi_{xyyy}-2\chi_{xzzy}\\&+\chi_{yyxy})\text{Cos}[\beta]^2\text{Sin}[\beta]-(\chi_{xzxx}+\chi_{zxxx})\text{Cos}[\beta]\text{Sin}[3\alpha]\text{Sin}[\beta]^2\\&+\chi_{zxyz}\text{Sin}[\beta]^3+\text{Cos}[3\alpha]\text{Cos}[\beta](-\chi_{yyzy}\text{Cos}[\beta]^2+(\chi_{yzyy}\\&+\chi_{zyyy})\text{Sin}[\beta]^2))\text{Sin}[\phi])))^2;\end{aligned}$$

For domain B:

$$\begin{aligned}SHG_{parallel,B} = &(-\chi_{yyzy}\text{Cos}[\beta]^4\text{Cos}[\phi]^3\text{Sin}[3\alpha]+\text{Cos}[\beta]^3\text{Cos}[\phi]^2((\chi_{xyxy}+2\chi_{xyyx}\\&-2\chi_{yzzy}-\chi_{zyzy})\text{Cos}[\phi]\text{Sin}[\beta]-3\chi_{xxzx}\text{Sin}[3\alpha]\text{Sin}[\phi])\\&+\text{Cos}[\beta]^2\text{Cos}[\phi]((2\chi_{yzyy}+\chi_{zyyy})\text{Cos}[\phi]^2\text{Sin}[3\alpha]\text{Sin}[\beta]^2\\&+\chi_{yyxy}\text{Cos}[\phi]\text{Sin}[\beta]\text{Sin}[\phi]+3\chi_{yyzy}\text{Sin}[3\alpha]\text{Sin}[\phi]^2)\\&+\text{Cos}[3\alpha](\text{Cos}[\beta]^2\text{Cos}[\phi]^3(\chi_{xxzx}\text{Cos}[\beta]^2-(2\chi_{xzxx}+\chi_{zxxx})\text{Sin}[\beta]^2)\\&+\text{Cos}[\beta]\text{Cos}[\phi]^2(-3\chi_{yyzy}\text{Cos}[\beta]^2+2(2\chi_{yzyy}+\chi_{zyyy})\text{Sin}[\beta]^2)\text{Sin}[\phi]\\&+\text{Cos}[\phi](-3\chi_{xxzx}\text{Cos}[\beta]^2+(2\chi_{xzxx}+\chi_{zxxx})\text{Sin}[\beta]^2)\text{Sin}[\phi]^2\\&+\chi_{yyzy}\text{Cos}[\beta]\text{Sin}[\phi]^3)-\text{Sin}[\beta]\text{Sin}[\phi]((\chi_{xzyz}+2\chi_{zxyz})\text{Cos}[\phi]^2\text{Sin}[\beta]^2\\&+\chi_{zyyy}\text{Cos}[\phi]\text{Sin}[3\alpha]\text{Sin}[\beta]\text{Sin}[\phi]-\chi_{yyxy}\text{Sin}[\phi]^2\\&+\chi_{yzyy}\text{Sin}[3\alpha]\text{Sin}[\beta]\text{Sin}[2\phi])+\text{Cos}[\beta]((\chi_{yzyz}+2\chi_{zyyz}\\&-\chi_{zzzz})\text{Cos}[\phi]^3\text{Sin}[\beta]^3+2(2\chi_{xzxx}+\chi_{zxxx})\text{Cos}[\phi]^2\text{Sin}[3\alpha]\text{Sin}[\beta]^2\text{Sin}[\phi]\\&+(\chi_{xyxy}-2\chi_{yzzy}-\chi_{zyzy})\text{Cos}[\phi]\text{Sin}[\beta]\text{Sin}[\phi]^2+\chi_{xxzx}\text{Sin}[3\alpha]\text{Sin}[\phi]^3\\&+\chi_{xyyx}\text{Sin}[\beta]\text{Sin}[\phi]\text{Sin}[2\phi]))^2;\end{aligned}$$

$$\begin{aligned}
SHG_{cross,B} = \frac{1}{16} &(4\chi_{xzyz}\cos[\phi]^3\sin[\beta]^3 - 4\chi_{yyzy}\cos[\beta]^4\cos[\phi]^2\sin[3\alpha]\sin[\phi]\\
&+ 4\cos[\phi]^2(2\chi_{yzyy}\sin[3\alpha]\sin[\beta]^2 - \chi_{xyyx}\sin[2\beta])\sin[\phi]\\
&- 4\cos[\phi]\sin[\beta](\chi_{yyxy} + 2\chi_{zxyz}\sin[\beta]^2)\sin[\phi]^2\\
&- 4\chi_{zyyy}\sin[3\alpha]\sin[\beta]^2\sin[\phi]^3 + \cos[\beta]^3(\chi_{xxzx}(\cos[\phi]\\
&+ 3\cos[3\phi])\sin[3\alpha] + 4(\chi_{xyxy} + 2\chi_{xyyx} - 2\chi_{yzzy}\\
&- \chi_{zyzy})\cos[\phi]^2\sin[\beta]\sin[\phi]) + \cos[3\alpha](\chi_{yyzy}\cos[\beta]^3(\cos[\phi]\\
&+ 3\cos[3\phi]) + 4\cos[\beta]\cos[\phi](\chi_{zyyy} - (2\chi_{yzyy} + \chi_{zyyy})\cos[2\phi])\sin[\beta]^2\\
&+ 4\cos[\phi]^2(\chi_{xxzx}\cos[\beta]^2(2 + \cos[\beta]^2) - (2\chi_{xzxx} + (2\chi_{xzxx}\\
&+ \chi_{zxxx})\cos[\beta]^2)\sin[\beta]^2)\sin[\phi] - 4\chi_{yyzy}\cos[\beta]\cos[\phi]\sin[\phi]^2\\
&+ 4(-\chi_{xxzx}\cos[\beta]^2 + \chi_{zxxx}\sin[\beta]^2)\sin[\phi]^3) + 2\cos[\beta]^2(2(\chi_{xyyy}\\
&- 2\chi_{xzzy})\cos[\phi]^3\sin[\beta] + (-4\chi_{yyzy} + 2\chi_{yzyy} + \chi_{zyyy} - (2\chi_{yzyy}\\
&+ \chi_{zyyy})\cos[2\beta])\cos[\phi]^2\sin[3\alpha]\sin[\phi] + 2\chi_{yyzy}\sin[3\alpha]\sin[\phi]^3 + (\chi_{xyyy}\\
&- 2\chi_{xzzy} + \chi_{yyxy})\sin[\beta]\sin[\phi]\sin[2\phi])\\
&+ 4\cos[\beta](-2\chi_{xzxx}\cos[\phi]^3\sin[3\alpha]\sin[\beta]^2 + \cos[\phi]^2\sin[\beta](2\chi_{yzzy}\\
&+ (\chi_{yzyz} + 2\chi_{zyyz} - \chi_{zzzz})\sin[\beta]^2)\sin[\phi] - \chi_{xxzx}\cos[\phi]\sin[3\alpha]\sin[\phi]^2\\
&+ \sin[\beta]\sin[\phi]((\chi_{xyxy} - \chi_{zyzy})\sin[\phi]^2 + (\chi_{xzxx}\\
&+ \chi_{zxxx})\sin[3\alpha]\sin[\beta]\sin[2\phi])))^2;
\end{aligned}$$

where $\alpha$ and $\beta$ are the cutting azimuth angle and the cutting polar angle, respectively.

d. **On the tilt-cut sample in the paraphase ($\bar{3}m$)**

$$\begin{aligned}
SHG_{parallel} = \Big(&-\chi_{xxzx}\cos[\beta]^4\cos[\phi]^3\sin[3\alpha]\\
&- (2\chi_{xzxx} + \chi_{zxxx})\cos[\phi]\sin[3\alpha]\sin[\beta]^2\sin[\phi]^2\\
&+ \cos[\beta]^3\cos[\phi]^2\Big((-\chi_{xyxy} - 2\chi_{xyyx} + 2\chi_{yzzy} + \chi_{zyzy})\cos[\phi]\sin[\beta]\\
&- 3\chi_{xxzx}\cos[3\alpha]\sin[\phi]\Big)\\
&+ \cos[\beta]^2\cos[\phi]\sin[3\alpha](\chi_{zxxx}\cos[\phi]^2\sin[\beta]^2 + 3\chi_{xxzx}\sin[\phi]^2)\\
&+ \cos[\beta]\Big(\cos[\phi]^3\sin[\beta]\big((-\chi_{yzyz} - 2\chi_{zyyz} + \chi_{zzzz})\sin[\beta]^2\\
&+ \chi_{xzxx}\sin[3\alpha]\sin[2\beta]\big) + 2(2\chi_{xzxx} + \chi_{zxxx})\cos[3\alpha]\cos[\phi]^2\sin[\beta]^2\sin[\phi]\\
&+ (-\chi_{xyxy} - 2\chi_{xyyx} + \chi_{zyzy})\cos[\phi]\sin[\beta]\sin[\phi]^2 + \chi_{xxzx}\cos[3\alpha]\sin[\phi]^3\\
&+ \chi_{yzzy}\sin[\beta]\sin[\phi]\sin[2\phi]\Big)\Big)^2
\end{aligned}$$

$$SHG_{cross} = \frac{1}{64}(\text{Cos}[3\alpha]\text{Cos}[\beta]((5\chi_{xxzx} - 4\chi_{xzxx} - 2\chi_{zxxx} + (3\chi_{xxzx} + 4\chi_{xzxx} + 2\chi_{zxxx})\text{Cos}[2\beta])\text{Cos}[3\phi] - 2(\chi_{xxzx} + 4\chi_{xzxx} - 2\chi_{zxxx})\text{Cos}[\phi]\text{Sin}[\beta]^2)$$
$$+ \text{Sin}[\phi](-8\chi_{xxzx}\text{Cos}[\beta]^2\text{Cos}[2\phi]\text{Sin}[3\alpha] + \text{Cos}[\phi]^2((-7\chi_{xxzx} + 10\chi_{xzxx} + \chi_{zxxx} - 8(\chi_{xxzx} + \chi_{zxxx})\text{Cos}[2\beta] - (\chi_{xxzx} + 2\chi_{xzxx} + \chi_{zxxx})\text{Cos}[4\beta])\text{Sin}[3\alpha] - 2(\chi_{xyxy} - 2\chi_{xyyx} + \chi_{yzyz} + 2\chi_{yzzy} + 2\chi_{zyyz} - \chi_{zyzy} - \chi_{zzzz})\text{Sin}[2\beta] + (-\chi_{xyxy} - 2\chi_{xyyx} + \chi_{yzyz} + 2\chi_{yzzy} + 2\chi_{zyyz} + \chi_{zyzy} - \chi_{zzzz})\text{Sin}[4\beta]) - 8\text{Sin}[\beta]((\chi_{xyxy} - \chi_{zyzy})\text{Cos}[\beta] + \chi_{zxxx}\text{Sin}[3\alpha]\text{Sin}[\beta])\text{Sin}[\phi]^2))^2$$

where $\alpha$ and $\beta$ are the cutting azimuth angle and the cutting polar angle, respectively.

The susceptibility tensors for domain A and B that lead to the functional forms under the $\bar{3}$ point group are related by the mirror operation along the *yz*-plane. During the fitting process, SHG RA patterns from both domain A and domain B in both the parallel and the cross channels are fitted together and yield robust fitting results.

## 2. Oblique SHG RA measurement on the unannealed normal-cut NiTiO$_3$

We confirm the bulk electric quadrupole (EQ) as the origin of the SHG signal and rule out the possibility of surface electric dipole (ED) contribution by performing oblique incidence SHG RA measurements on the normal-cut NiTiO$_3$ sample. The simulated SHG RA signal as a function of the incident polarization angle $\phi$ in the S-P channel is given by

$$SHG_{sp,3} = \chi_{zxx}^2\text{Sin}[\theta]^2 + \text{Cos}[\theta]^2(\chi_{xxx}\text{Cos}[3\phi] + \chi_{yyy}\text{Sin}[3\phi])^2$$

for the surface ED (point group 3), and

$$SHG_{sp,\bar{3}} = \text{Cos}[\theta]^2(-\chi_{xyxy}\text{Sin}[\theta] + \text{Cos}[\theta](\chi_{xxzx}\text{Cos}[3\phi] + \chi_{yyzy}\text{Sin}[3\phi]))^2 + \text{Sin}[\theta]^2(-\chi_{zyzy}\text{Cos}[\theta] + \text{Sin}[\theta](\chi_{zxxx}\text{Cos}[3\phi] + \chi_{zyyy}\text{Sin}[3\phi]))^2$$

for the bulk EQ (point group $\bar{3}$), where $\chi_{ijk}$ and $\chi_{ijkl}$ are the susceptibility tensors for the ED and EQ SHG process; $\theta$ and $\phi$ are the incident polar angle and the polarization angle of the incident light. Figure S1A and B plot the simulated oblique incidence SHG RA patterns in the S-P channel for surface ED and bulk EQ, respectively. We see that the surface ED SHG RA pattern has six even lobes in the S-P channel whereas the bulk EQ SHG RA shows six lobes with alternating amplitudes. Hence, the oblique incidence SHG RA technique can be used to distinguish the origin of the SHG signal. Figure S1C shows the oblique incidence SHG RA pattern in the S-P channel (markers) from the normal-cut sample, which clearly shows alternating lobe amplitudes and is consistent with the bulk EQ SHG RA fit (solid curve). This confirms that our SHG signal is primarily contributed by bulk EQ instead of surface ED.

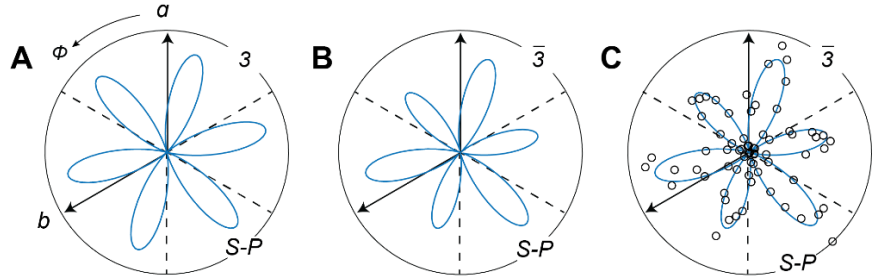

**FIG. S1.** Simulated oblique incidence SHG RA pattern in the S-P channel from (**A**) surface ED (point group 3) and (**B**) bulk EQ (point group $\bar{3}$). (**C**) Oblique incidence SHG RA (markers) fitted by bulk EQ simulation (solid line).

## 3. Normal incidence EQ SHG RA of the unannealed normal-cut NiTiO$_3$

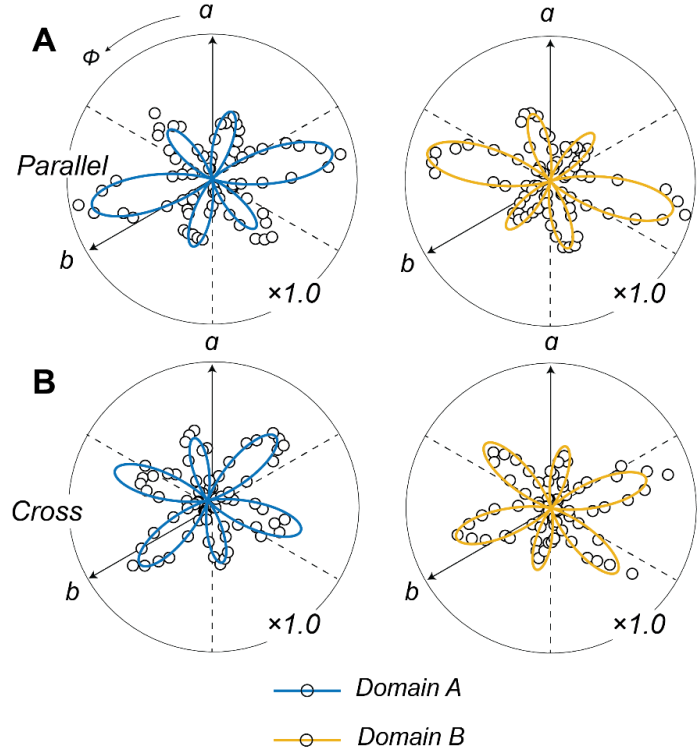

**FIG. S2.** EQ SHG RA of the unannealed normal-cut NiTiO$_3$ in (**A**) the parallel channel and (**B**) the cross channel. Data (markers) are fitted by the functional form derived by group theory (solid lines). Mirrors of the paraphase are indicated by the dash lines. Numbers at the bottom right of the plots indicate the scale of the plot.

The surface normal of the normal-cut NiTiO$_3$ sample is off from the crystal $c$-axis by $\theta_1 = 2 \pm 1°$, determined by the fitting of the EQ SHG RA data shown in Fig. S2. Given this small deviating angle, we normalize the six lobes of the EQ SHG RA patterns and plot them in Fig. 2 of the main text.

## 4. SHG signal contrast between the normal-cut and the tilt-cut sample

Comparing to the normal-cut sample, the tilt-cut sample increases the anisotropy of the SHG RA patterns, making it possible to nicely see the contrast between two FR domains over a wide polarization angle range. Specifically, the SHG RA patterns from the tilt-cut sample only have two-fold rotational symmetry (Figure S3A shows the parallel channel SHG RA data) whereas those from the normal-cut sample have the six-fold rotation symmetry (Figure S3B shows the parallel channel SHG RA data). Besides, it further provides a greater SHG intensity difference between the two FR domain states. Figure S3C shows the SHG signal difference between domain A and domain B for the tilt-cut (blue) and the normal-cut (red) samples under the same measurement conditions. It is clearly seen that the signal difference between the two FR domains for the tilt-cut sample is much larger than that of the normal-cut sample. This large signal difference provides a sharper contrast of the FR domains in the SHG maps, as you may compare Fig. 3A and 3B in the main text and Fig. S6 in the Supplementary Materials.

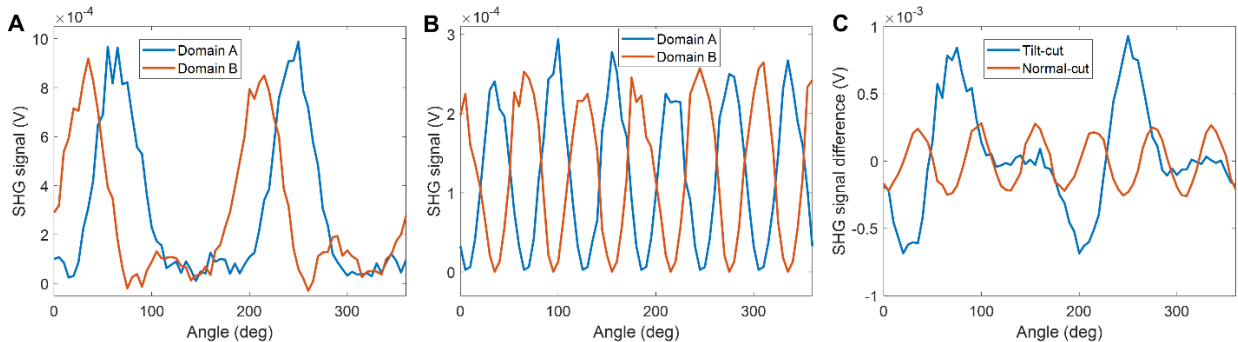

**FIG. S3.** SHG RA signal as a function of the incident light polarization angle of domain A (blue) and domain B (red) from (**A**) tilt-cut sample and (**B**) normal-cut sample. (**C**) SHG RA signal difference between domain A and domain B from the tilt-cut sample (blue) and the normal-cut sample (red).

## 5. FR domain maps color scale determination

The determination of the FR domain mapping color scale shown in Fig. 3A and 3B is described by the following procedure. First, we extract line-cuts from both the $\phi = -17°$ and the $\phi = 17°$ domain mappings at the same location (dash lines in Fig. S4B and S4C), and plot them in Fig. S4A. We then choose the cross-points between the two curves to be the signal level of the domain wall (dash line in Fig. S4A) and set it to be color grey. The upper bound and lower bound colors are set to be blue and orange at the average signal levels of the two domains respectively.

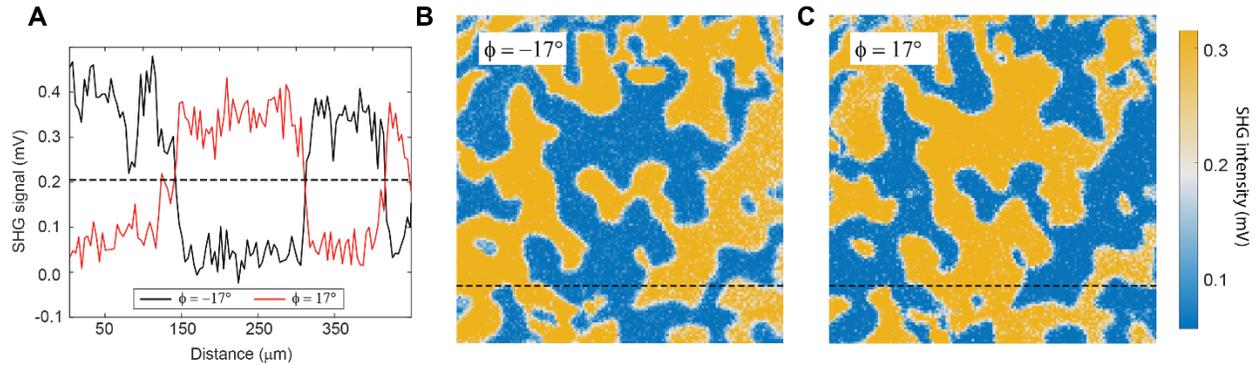

**FIG. S4.** (**A**) Line-cuts extracted at the same location from (**B**) $\phi = -17°$ and (**C**) $\phi = 17°$ EQ SHG FR domain maps. The dash line corresponds to the signal level of the domain walls. (**B**) EQ SHG FR domain mapping performed at $\phi = -17°$ and (**C**) $\phi = 17°$ after updating the color scale. The dash lines correspond to the locations where the line-cuts of the SHG signal are taken and plotted in (**A**).

## 6. Exceptionally large SHG signals from NiO defects

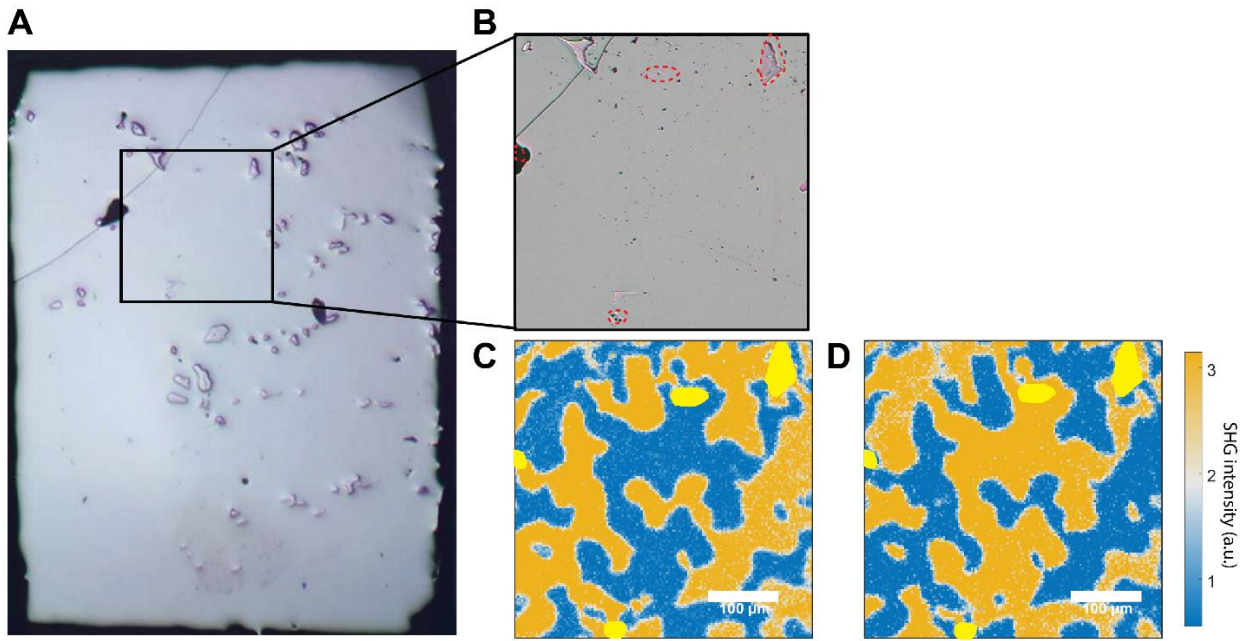

**FIG. S5.** (**A**) Optical image of the NiTiO$_3$ sample that was used for EQ SHG measurement. (**B**) Zoomed-in area where the SHG scanning is performed. Regions with exceptionally large SHG signals correspond to NiO and are circled by the red dash curves. (**C**) and (**D**) EQ SHG scan maps from the region shown in (**B**) under the polarization (**C**) $\phi = 17°$ and (**D**) $\phi = -17°$. Regions with exceptionally large SHG signals are highlighted by the yellow color.

## 7. EQ SHG scanning microscopy of the unannealed normal-cut NiTiO$_3$

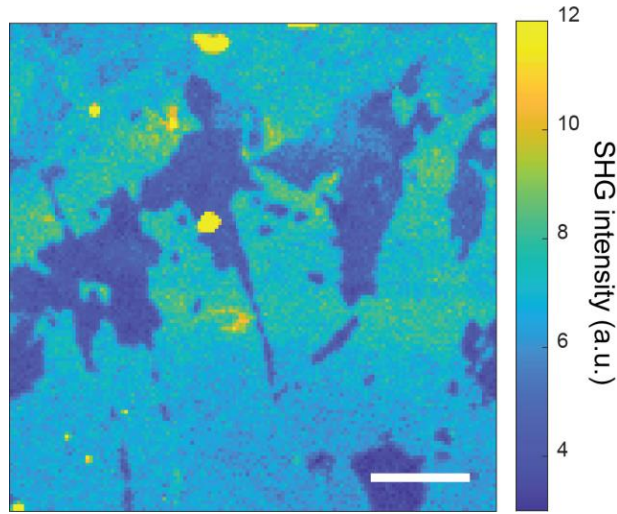

**FIG. S6.** EQ SHG domain mapping of the unannealed normal-cut NiTiO$_3$. Scale bar corresponds to 100 μm. The domain size is much larger than that of the annealed sample.

## 8. Radius of curvature along the domain walls of the annealed and quenched tilt-cut NiTiO$_3$

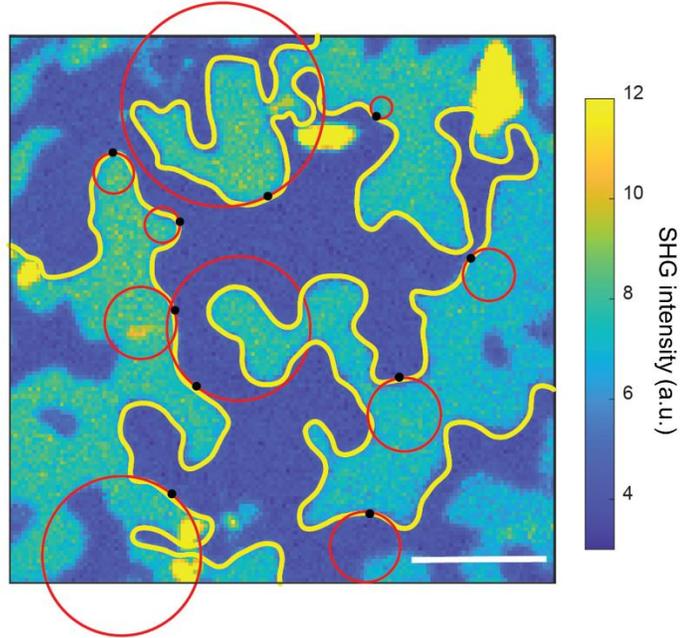

**FIG. S7.** EQ SHG domain mapping from the annealed and quenched tilt-cut sample with domain walls highlighted. Circles with the calculated radius of curvature as the radius are illustrated at selected points on the boundary. Black dots indicate the position where the radius of curvature is calculated. Scale bar corresponds to 100 μm.

The radius of curvature is calculated at each point on the domain walls by the formula

$$R = \left| \frac{(1+y')^{\frac{3}{2}}}{y''} \right|,$$

where $y' = \frac{\Delta y}{\Delta x}$ and $y'' = \frac{\Delta y\prime}{\Delta x}$. Circles with the radius of curvature calculated as the radius at selected positions on the FR domain walls are illustrated in Fig. S7.

# 9. Discussion of destructive interference at the domain boundary

We have carefully considered the possibility of destructive-interference-induced SHG intensity suppression at domain boundaries. In this section, we work along the following 3 steps to rule it out.

**I) We have designed our experiment differently from the literature**

In literature, the wide field SHG imaging technique has been used to visualize domain walls as dark lines (*i.e.*, suppressed SHG intensity as compared to domains) (*27, 29, 50, 51*). In their settings, the following two conditions are fulfilled: a) the domain states on the two sides of the domain wall have opposite SHG susceptibility tensors, i.e. $\chi_{domain\,I} = -\chi_{domain\,II}$; b) the focusing objective, sample, imaging lens, and the detector form a nice imaging system that maps the light from the sample to the detector point by point up to the diffraction-limited resolution. These two conditions make the light from the two domains around the domain wall within the diffraction limit to destructively interfere when they arrive at the same location on the detector site, leading to the (fully) suppressed SHG signal at the domain boundaries.

In our experiment, we used scanning SHG microscopy technique where the focused laser beam raster-scan across the sample and the reflected SHG signal is sent to a PMT detector. For our system and experimental setups, they are different from the literature for both conditions: a) our domain states are related by a mirror operation, i.e., $\chi_{domain\,I} = m[\chi_{domain\,II}]$, that does not lead to the opposite sign of the whole tensor, and hence the full destructive interference, as the literature does; b) we deliberately put the PMT detector significantly off-focus from the lens in front of it so as to ruin the imaging condition between the sample and the detector as well as prevent the light across the domain walls from arriving at the same physical spot on the detector.

Due to the difference in both a) and b) conditions, it is unlikely that our slightly suppressed intensity at the domain walls originates from the destructive interference as the literature does.

**II) We have performed a control experiment to rule out destructive interference in our scanning SHG map**

We choose $Ni_3TeO_6$ as our test candidate which has two domain states with opposite SHG susceptibility tensors, fulfilling condition a) in literature. We have performed the wide field SHG imaging in the very same way as described in literature, fulfilling condition b), and also our scanning SHG microscopy with the off-focus setting between the detector and its collecting lens, violating condition b). We show in Fig. S8A and S8B that dark lines (*i.e.*, a lower SHG signal level) appear at the boundaries between domains in the wide field SHG imaging graph whereas in the SHG scanning map shown in Fig. S8C, the domain boundaries are bright, *i.e.*, at a higher signal level than within domains. The contrasting behavior we observed here convinces us that our scanning SHG microscopy removes the destructive interference effect that is seen in the wide field SHG imaging.

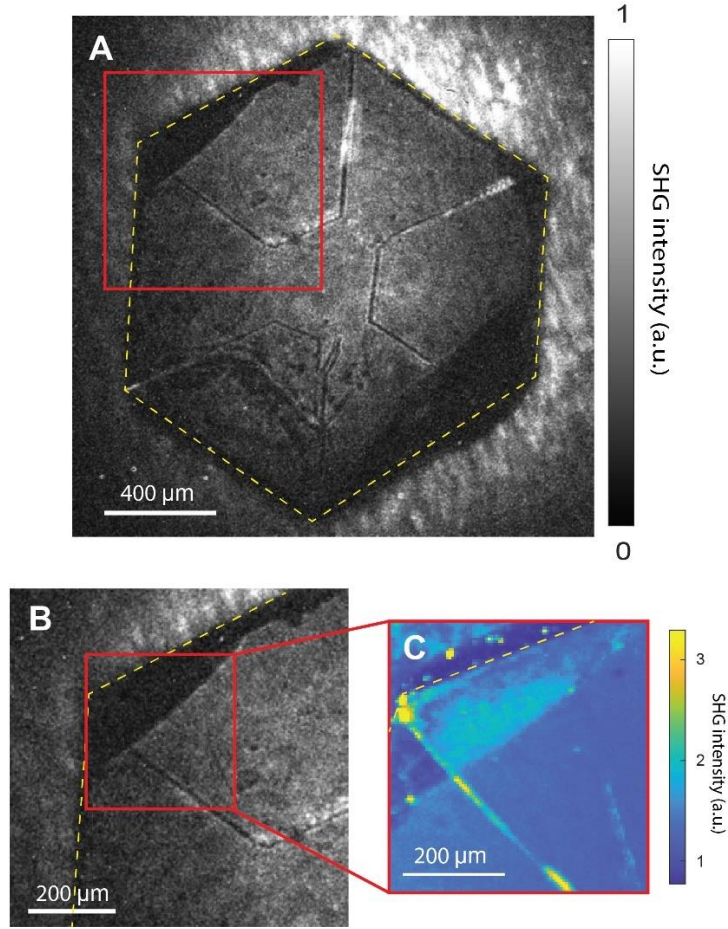

**FIG. S8.** (**A** and **B**) Wide field SHG image of $Ni_3TeO_6$. Red square in (**A**) corresponds to the region shown in (**B**). Red square in (**B**) corresponds to the region where scanning SHG was performed. The domain boundary appears to be dark lines in the wide field SHG image. (**C**) Scanning SHG microscopy of $Ni_3TeO_6$ taken within the red square shown in (**B**). The domain boundary appears to be bright lines instead. Yellow dash lines serve as guide to the eyes of the sample edges.

### III) We have performed scanning electron microscopy (SEM) measurements to check the domain wall width

We have further performed SEM measurements to check the domain wall width. Figure S9A shows the secondary electron SEM image at 5.0 kV that reveals the domain walls as dark lines, coinciding with the domain boundaries shown in the SHG map (Fig. S9B). Fig S9C shows the SEM image of a zoomed-in area that is indicated by the red square in Fig. S9A. By extracting twenty linecuts across the domain wall in Fig. S9C and averaging over them (yellow lines in Fig. S9C are representative line-cuts), we construct the line profile across the domain wall and show it in Fig. S9D, from which we see that the domain wall width lies in the range from 500 nm to 1 µm, consistent with the value estimated by our SHG measurements.

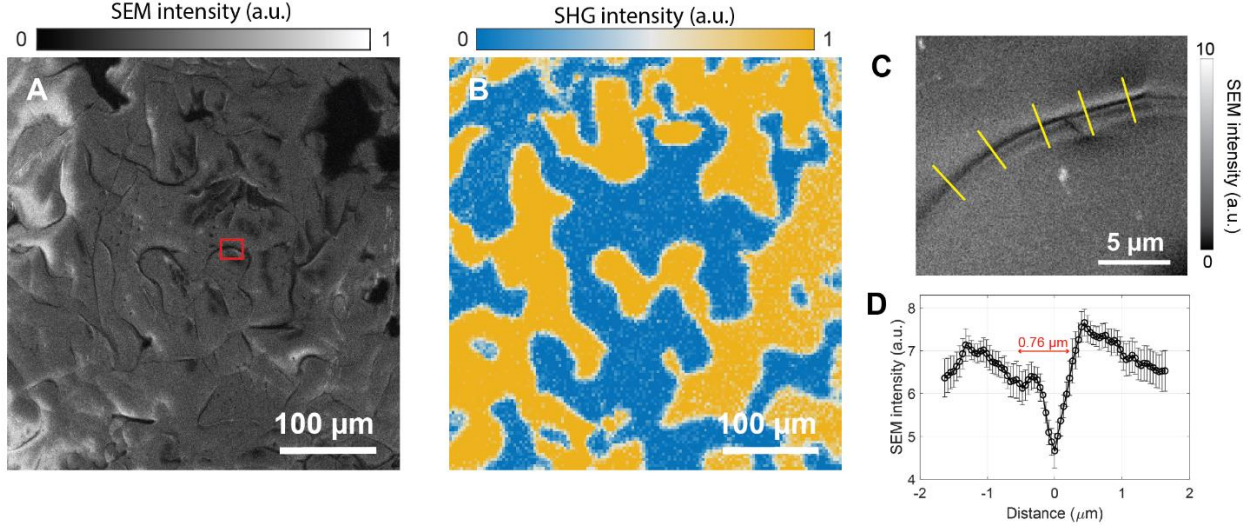

**FIG. S9.** (**A**) The secondary electron SEM image at 5.0 kV that reveals the domain wall. Red square corresponds to the region shown in (**C**). (**B**) SHG map of the same region as (**A**) performed at $\phi = -17°$. (**C**) SEM image of the zoomed-in section indicated by the red square in (**A**). Yellow lines are representative line-cuts that are used to construct (**D**). (**D**) Line profile across the domain wall in (**C**) calculated by averaging over twenty line-cuts across the domain wall shown in (**C**). Length that corresponds to 0.76 µm is indicated by the red arrow. The domain wall width lies in the range from 500 nm to 1 µm.

With the efforts in I-III) above, we are confident that the lower SHG signal at the interfaces between domains in NiTiO$_3$ in our SHG map mainly originates from the restoration of mirror symmetry at the domain walls, instead of from the destructive interference effect. Based on this, we performed an estimate of the domain wall width by computing the measured SHG intensity at the dip of the linecut in Fig. 4C with the weighted sum of SHG intensities from 1) domain A; 2) domain B; and 3) domain wall, where domains A and B have the same SHG intensity and domain wall has zero SHG intensity at the selected polarization angle. The beam size in our study here is 15 $\mu m$ FWHM which is consistent with the width of the dip in Fig. 4C. The calculated domain wall width $w = 0.76 \pm 0.20 \, \mu m$ is based on the depth of the dip and the estimation model mentioned above.

## 10. Estimation of the domain wall width

Assuming the excitation laser beam has an intensity profile of

$$P(x,y) = A e^{-\frac{x^2}{2\sigma^2}} e^{-\frac{y^2}{2\sigma^2}}$$

where A is the laser intensity at the center of the beam and $\sigma$ is the standard deviation of the Gaussian beam profile, which is related to the full width at half maximum (FWHM = 15 $\mu m$) of the laser beam by

$$FWHM \approx 2.355 \, \sigma.$$

The SHG generated will be proportional to the square of the incident power, namely,

$$SHG(x,y) \propto P(x,y)^2 = A^2 e^{-\frac{x^2}{\sigma^2}} e^{-\frac{y^2}{\sigma^2}}.$$

The total SHG generated inside the beam region without the domain wall will then be

$$SHG_{total} = \int_{-\infty}^{\infty} dx \int_{-\infty}^{\infty} dy\ P(x,y) = \pi A^2 \sigma^2.$$

Now, assume that there is a domain wall region with a width $w$ at the center of the laser beam. The SHG that corresponds to this region is then

$$SHG_{wall} = \int_{-w/2}^{w/2} dx \int_{-\infty}^{\infty} dy\ P(x,y) = \pi A^2 \sigma^2\ \text{erf}\left(\frac{w}{2\sigma}\right),$$

which corresponds to the $7.0 \pm 1.5\%$ drop of the SHG signal shown in Fig. 4C

$$\frac{SHG_{wall}}{SHG_{total}} = 7.0 \pm 1.5\%.$$

Solving for $w$ we have $w = 0.76 \pm 0.20\ \mu m$.